\documentclass[a4paper,11pt]{article}
\usepackage{jheppub} 
\usepackage{lineno}
 \usepackage{graphics}
 \usepackage{float}
 \usepackage{tikz-cd} 
\usepackage{url}
\usepackage{hyperref}
\usepackage{float}
\usepackage{standalone}
\usepackage{bm}
\usepackage{quiver}
\usepackage{cleveref}
\usepackage{subcaption}
\usepackage{graphicx} 
\usepackage{array}
\usepackage{paralist}
\usepackage{amsmath}
\usepackage{physics}
\usepackage{float}
\usepackage{xcolor}
\usepackage{amsfonts}
\usepackage{amssymb}
\usepackage{verbatim}
 \usepackage{indentfirst}
\usepackage{tikz}
\usepackage{soul}
\usepackage{pifont}
\usepackage{tikz-cd}
\tikzcdset{
	arrow style=tikz,
	diagrams={>={Straight Barb[scale=0.6]}},
    every arrow/.append style={line width=0.4pt}
}
\usetikzlibrary{calc}
\tikzset{curve/.style={settings={#1},to path={(\tikztostart)
    .. controls ($(\tikztostart)!\pv{pos}!(\tikztotarget)!\pv{height}!270:(\tikztotarget)$)
    and ($(\tikztostart)!1-\pv{pos}!(\tikztotarget)!\pv{height}!270:(\tikztotarget)$)
    .. (\tikztotarget)\tikztonodes}},
    settings/.code={\tikzset{quiver/.cd,#1}
        \def\pv##1{\pgfkeysvalueof{/tikz/quiver/##1}}},
    quiver/.cd,pos/.initial=0.25,height/.initial=0}

\usepackage{caption, subcaption}
 
 \usepackage{tikz}
\usepackage{multirow} 
\usepackage{makecell}
\usepackage{longtable}

\def\diag{\mathop{\rm diag}\nolimits}

\def\rank{\mathop{\rm rank}}

\def\beq{\begin{eqnarray}}
\def\eeq{\end{eqnarray}}
\def\beq#1\eeq{\begin{align}#1\end{align}}

\title{Three dimensional quotient singularity and 4d $\mathcal{N}=1$ AdS/CFT correspondence}
\author{Yuanyuan Fang,}
\author{Jing Feng}
\author{and Dan Xie}

\affiliation{Department of Mathematics, Tsinghua University, Beijing, 100084, China}

\abstract{We systematically study the AdS/CFT correspondence induced by D3 branes probing three dimensional Gorenstein quotient singularity $\mathbb{C}^3/G$. The field theory is given by the McKay quiver, which has a vanishing NSVZ beta function assuming that all the chiral fields have the $U(1)_R$ charge $\frac{2}{3}$. Various physical quantities such as quiver Hilbert series, superconformal index, central charges, etc are computed, which match exactly with those computed using the singularity. We also study the relevant deformation of those theories and find the dual geometry, therefore generate many new interesting AdS/CFT pairs. The quiver gauge theory defined using finite subgroups of $SO(3)$ group  has some interesting features, for example, its Seiberg duality behavior is quite interesting. }

\begin{document} 
\maketitle

\flushbottom

\section{Introduction}
One of the best way to get AdS/CFT pairs is to use D branes or M2/M5 branes 
to probe a singularity $X$ of manifold with special holonomy \cite{Morrison:1998cs}, and one gets the following dual pair:
\begin{equation*}
{\cal T} \longleftrightarrow  AdS_{d+1} \times L\,.
\end{equation*}
The superconformal field theory (SCFT) ${\cal T}$ is realized as the IR  field theories living on the branes, and the internal geometry $L$ on the string/M theory side is given by the link of the singularity $X$ \cite{looijenga1984isolated} \footnote{The link is often defined as the intersection of $X$ with a sphere with radius one.}. When $X$ is smooth, one gets the maximal supersymmetric dual pairs studied by Maldacena \cite{Maldacena:1997re}. 
An early example of using singularity was the dual pair studied by 
Klebanov and Witten \cite{Klebanov:1998hh} where they used three dimensional confiold singularity.

The space of $X$ which would give rise to the dual pair is very large if one is interested in the theory with the minimal number of supersymmetry. For example, it was conjectured in \cite{Xie:2019qmw} that one can use three dimensional Gorenstein canonical singularity $X$ \cite{reid46young} to get 4d $\mathcal{N}=1$ AdS/CFT pair, and the space of such 3d singularities is 
very large, see \cite{Xie:2019qmw}. Further studies of those dual pairs would be quite valuable for us to better understand 
gauge/gravity duality. However, we face two major problems in these studies: a): First, not every 3d canonical singularities would give us an AdS/CFT pair, as there are obstructions \cite{gauntlett2007obstructions,collins2019sasaki} to the existence of a Ricci flat conic metric on $X$, which is a necessary and sufficient condition for the existence of AdS/CFT pair; b): Even if $X$ would give rise to a AdS/CFT pair, the field theory is quite difficult to define and most of them do not admit 
a conventional Lagrangian theory (i.e. the SCFT is not realized as the IR limit of a usual non-abelian gauge theory).

When $X$ is a three dimensional toric \footnote{A 3d toric singularity admits three dimensional automorphism group.} singularity, it was shown in \cite{futaki2009transverse} that $X$ is indeed stable and one can find the gauge theory by using the dimer technique \cite{franco2006brane,franco2006gauge,yamazaki2008brane}. 

Another class of stable singularities is the so-called Gorenstein quotient singularities $\mathbb{C}^3/G$, where $G$ is 
a finite subgroup of $SU(3)$. Such singularities were classified in \cite{Yau1993GorensteinQS}, see Table \ref{fgsu3}, and the corresponding link $L$ is an orbifold  $S^5/G$ \footnote{In general, the space $S^5/G$ is not a smooth manifold.}. The SCFT also admits a simple description by using the representation theory of the group $G$, namely the so-called McKay quiver \cite{ito1996mckay,bridgeland2001mckay,gomi2000hilbert,gomi2004coinvariant}. So one can always get a nice AdS/CFT pair (see \cite{Lawrence:1998ja, Kachru:1998ys,Douglas:1997de,Greene:1998vz,Hanany:1998sd} for the early studies). Comparing with toric theories, these pairs are often simpler both on the geometry side and field theory side, i.e. many singularities $X$ are given by hypersurface and the field theory might have just a few simple gauge groups. The McKay quiver 
also has  quite nice features: the one loop beta function all vanishes (finite theory) assuming all the chiral fields do not receive anomalous dimension, and so it is similar to $\mathcal{N}=4$
SYM and $\mathcal{N}=2$ affine ADE quiver gauge theory. So those AdS/CFT provide an excellent class of examples from which one can learn deep lessons about field theory and string theory.

In this paper, we will perform a systematical study of those AdS/CFT pairs by using various  mathematical and physical tools newly developed in the last few years: 
\begin{enumerate}
\item One can compute a so-called quiver matrix Hilbert series \cite{ginzburg2006calabi,bocklandt2008graded,eager2013can,eager2014equivalence}, from which one can extract 
a Hilbert series $H_{00}(t)$. $H_{00}(t)$ can be used to extract the dual geometry by identifying it as the Hilbert series of the singularity $X$. 
\item The superconformal index has a quite simple form in the large $N$ limit \cite{ginzburg2006calabi,Gadde:2010en}, and 
one can get the exact formula for those theories, from which the information of protected spectrum is derived.
\item The Seiberg duality \cite{Seiberg:1994pq} behaviors of the McKay quivers are richer, and the dual theory is no longer finite (the chiral fields receive anomalous dimensions). Geometrically, Seiberg duality is related to the flop of the crepant resolution of the singularity. Given the simple nature of the geometry, 
those pairs might teach us more lessons about Seiberg duality.
\item We also study the relevant deformation of these theories, and find the dual geometry by 
computing the quiver Hilbert series. In particular, the mass deformed $\mathcal{N}=2$ affine ADE quiver is studied, where the dual geometry is proven to be:
\begin{equation*}
f_{ADE}(x,y,z)+w^h=0\,.
\end{equation*}
Here $f_{ADE}$ is the ADE singularity and $h$ is the Coxeter number. This is the generalization of 
the conifold example \cite{Klebanov:1998hh} (see also \cite{Gubser:1998ia, cachazo2001geometric} for the early discussion). The dual geometry is nicer as 
it has only isolated singularity so the link $L$ is smooth; and various geometric properties of $L$ such as homology groups are computed. We also compute the Hilbert series and index for those theories. Deformed theory for theories defined using finite subgroup of $SO(3)$ group is also studied, where the dual geometry is also written down.
\end{enumerate}

This paper is organized as follows: section \ref{finitesubgpsu3} reviews the classification of finite subgroup $G$ of $SU(3)$
and how to get a McKay quiver from the representation theory of $G$; section \ref{adscftcor} reviews how to compute various physical quantities from the field theory and geometry side; section \ref{finitesubgpsu2} studies the theory when $G$ is a subgroup of $SU(2)$; section \ref{fintesubgpso3} studies the theory when $G$ is a finite subgroup of $SO(3)$; Finally a conclusion is given in section \ref{conclusion}. Appendix studies other sporadic theories when $G$ is other finite subgroups of $SU(3)$.

\section{Finite subgroups of $SU(3)$ and McKay quiver}\label{finitesubgpsu3}
\subsection{Finite subgroups of $SU(3)$}
The finite subgroups of $SU(3)$ are classified in \cite{Yau1993GorensteinQS,Ludl:2011gn}, see the list in Table \ref{fgsu3}.
Here are some comments:
\begin{enumerate}
    \item  $\mathbb{C}^3/G$ has an isolated singularity at the origin if and only if $G$ is an abelian group
    and $1$ is not an eigenvalue for every element of $G$. 
    \item Finite groups of $SU(2)$ are contained in type $B$, and the elements of $G$ take the form
    \begin{equation}
   g=\left(\begin{array}{ccc}
   1&0\\
   0& g^{'} 
   \end{array}\right)\,.
    \end{equation}
       Here $g^{'}$ is the element of finite subgroup of $SU(2)$. They have an ADE classification.
    \item Finite groups of $SO(3)$ are identified as follows:
    \begin{enumerate}
        \item The cyclic group takes the form $A(n,1)$. 
        \item The dihedral group is contained in the $B$ series.
        \item The tetrahedral group is identified as $\Delta(12)$.
        \item The octahedral group is isomorphic to  $\Delta(24)$.
        \item The icosahedral group is isomorphic to $H$.
    \end{enumerate}
\end{enumerate}

\begin{table}
\begin{center}
\begin{tabular}{|c|c|} \hline 
Group & Order \\ \hline 
$A(m,n)\sim \mathbb{Z}_n\times \mathbb{Z}_m$,~$n$ divides $m$ & $mn$ \\ \hline
$B$~Finite subgroups of $U(2)$ & No general formulas \\ \hline
$C(n,a,b)$ & No general formulas \\ \hline
$D(n,a,b; d,r,s)$ & No general formulas \\ \hline
$\Delta(3n^2) \sim C(n,0,1),~n\geq 2$ & $3n^2$ \\ \hline
$\Delta(6n^2) \sim D(n,0,1;2,1,1), n\geq 2$ & $6n^2$ \\ \hline
$T_n \sim C(n,1,a),~(1+a+a^2)mod~n=0$ & $3n$ \\ \hline
$E$&  108  \\ \hline
$F$&  216  \\ \hline
$G$&  648  \\ \hline
$H$&  60  \\ \hline
$I$&  168  \\ \hline
$J$&  180  \\ \hline
$K$&  504  \\ \hline
$L$&  1080  \\ \hline
\end{tabular}
\end{center}
\caption{Finite groups of $SU(3)$.}
\label{fgsu3}
\end{table}

\subsection{McKay quiver}

The quiver gauge theory associated with $N$ $D3$ branes probing the singularity $\mathbb{C}^3/G$ can be derived using
the McKay correspondence \cite{ito1996mckay}.  Let's denote the irreducible representation of $G$ by $\rho_i$ whose dimension is $d_i$. The McKay quiver is found as follows:
\begin{enumerate}
    \item First the quiver nodes are identified with the irreducible representations of $G$: for an irreducible representation, assign a quiver node $Q_i$ whose gauge group is $SU(Nd_i)$.
    \item The number of quiver arrows from $Q_i$ to $Q_j$ is found from the following tensor product decomposition
    \begin{equation*}
\rho_i \otimes \rho =\oplus_j a_{ij} \rho_j\,.
        \end{equation*}
        Here $\rho$ is the natural representation of $G$: It is the three-dimensional representation on which $G$ acts naturally and could be reducible or irreducible.
        The number of arrows is just $a_{ij}$, and they point from quiver node $i$ to node $j$. We might also have adjoint chirals on $i$-th quiver node if $a_{ii}$ is nonzero.
        \item The superpotential can often be determined by requiring the gauge invariance, conformal invariance, and the total $R$ charge to be $2$ for each term in the superpotential. For our theory, the $U(1)_R$ charges for the chiral fields are all free ($\frac{2}{3}$), so the marginal interaction (total $R$ charge $2$) comes from oriented triangle (required by gauge invariance) in the quiver diagram. We simply add all possible such cubic terms to the superpotential.
\end{enumerate}

\begin{figure}[h]
    \centering
    \begin{tikzcd}[row sep=tiny, column sep=tiny]
				&[+20pt]&&[+20pt]&&3_2\ar[dl,bend left=10,""]\ar[dr,bend left=10,""]&\\[+30pt]
				1_0\ar[rr,bend left=10,""]&&3_1\ar[rr,bend left=10,""]\ar[ll,bend left=10,""]\ar[loop,out=290,in=250,looseness=9,""]&&5\ar[ll,bend left=10,""]\ar[rr,bend left=10,""]\ar[ur,bend left=10,""]\ar[loop,out=290,in=250,looseness=12,""]&&4\ar[ll,bend left=10,""]\ar[ul,bend left=10,""]\ar[loop,out=290,in=250,looseness=12,""]\\
			\end{tikzcd}
    \caption{McKay quiver corresponding to group $H$.}
    \label{quiverh}
\end{figure}
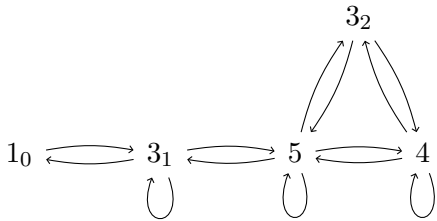

\textbf{Example}: Let's look at the group $H$, and the required representation theory data is given in \cite{gomi_nakamura_shinoda_2004}. There are a total of 5 irreducible representations with label $1_0, 3_1, 3_2, 4, 5$, where the number of the label denotes the dimension of the representation. The tensor product is given by
\begin{align*}
&1_0\otimes \rho=3_1,~~3_1\otimes \rho=1_0+3_1+5,~~3_2\otimes \rho=4+5, \nonumber \\
& 4\otimes \rho=3_2+4+5,~~5\otimes \rho=3_1+3_2+4+5\,.
\end{align*}
So there are a total of $5$ quiver nodes with rank $(1,3,3,4,5)$ for a single D3 brane probe. The number of arrows is
\begin{align*}
    &a_{1_03_1}=1,~~a_{3_1 1_0}=1,~~a_{3_13_1}=1,~~a_{3_15}=1,~~a_{3_2 4}=1,~~a_{3_25}=1, \nonumber\\
    & a_{43_2}=1,~~a_{44}=1,~~a_{45}=1,~~a_{53_1}=1,~~a_{53_2}=1,~~a_{54}=1,~~a_{55}=1\,.
\end{align*}
The resulting quiver gauge theory is shown in the Figure \ref{quiverh}. For $N$ D3 branes, one just changes the gauge group ranks to $(N,3N,3N,4N,5N)$.

 \subsubsection{General properties of McKay quiver}\label{mkproperty}
Let's discuss some general facts about the McKay quiver. We will show that the quiver gives rise to a 
4d $\mathcal{N}=1$ theory which does not have gauge anomaly, and the one loop beta function is zero
if all the chiral fields have the $U(1)_R$ charge $\frac{2}{3}$ (no anomalous dimension).

First, let's take a single D3 brane probe, so the McKay quiver has the dimension vectors $d_1, d_2\ldots, $
corresponding to the dimension of the irreducible representation of finite group $G$.
There is always a node corresponding to the trivial representation, and its rank is just $1$.

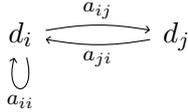
\begin{figure}[H]
 	\centering
 	\begin{tikzcd}[row sep=large,column sep=large]
 		d_i\ar[r,bend left=10,"a_{ij}"]\ar[loop,out=285,in=255,looseness=9,"a_{ii}"]& d_j\ar[l,bend left=10,"a_{ji}"]
 	\end{tikzcd}
 	\caption{The $i$th quiver node (with rank $d_i$) has $a_{ij}$ chiral fields coming out of node, and $a_{ji}$ chiral fields coming into the node, and $a_{ii}$ adjoint chirals.}
 	\label{exquiver}
 \end{figure}

We now show that the one loop beta function of McKay quiver is zero.  
Given a finite subgroup $G$ of $SU(3)$ group, one get a quiver by looking at the following branching rule 
for the irreducible representations
\begin{equation*}
\rho_i\otimes \rho =\oplus_j a_{ij} \rho_j,
\end{equation*}
where $a_{ij}$ gives the number of arrows between $i$th and $j$th quiver node. One defines a matrix $B$ as follows
\begin{equation*}
B_{ij}=3\delta_{ij}-a_{ij}\,.
\end{equation*}
We then define a matrix 
\begin{equation*}
A=B+B^t\,.
\end{equation*}
Let's now multiply $A$ by the dimension vector $d^T=[d_1, d_2, \ldots]$, and so 
\begin{equation}
(Ad)_i=\sum_j (6\delta_{ij}-a_{ij}-a_{ji})d_j\,.
\label{oneloop}
\end{equation}
We now show that the $i$th entry of $Ad$ is proportional to the one-loop beta function of the $i$th gauge group, assuming the $U(1)_R$ charge of all the chiral fields are $\frac{2}{3}$ (the R charge of a free chiral). The reason is the following: 
as shown in Figure \ref{exquiver}, for the $i$th gauge group with rank $d_i$, there are $d_j a_{ji}$ bi-fundamental chiral fields coming into it, 
 $a_{ij}d_j$ number of bi-fundamental chiral fields coming out of it, and $a_{ii}$ number of adjoint chiral fields, so the one loop NSVZ beta function (if all the chiral fields has $R$ charge $\frac{2}{3}$) is 
 \begin{align*}
& \beta_1=d_i+\sum_{j\neq i}[\frac{1}{2}(a_{ji}+a_{ij})d_j(\frac{2}{3}-1)]+a_{ii}d_i(\frac{2}{3}-1) \nonumber\\
& =\frac{1}{6}[6d_i-\sum_{j\neq i}(a_{ij}+a_{ji})d_j-2a_{ii} d_i] \nonumber\\
&=\frac{1}{6} \sum_j (6\delta_{ij}-a_{ij}-a_{ji})d_j\,.
 \end{align*}
Here we used the fact that the index for the fundamental representation is $\frac{1}{2}$, and that for the adjoint representation is $d_i$. Comparing with (\ref{oneloop}), we see that $(Ad)_i$ is just the one loop $\beta$ function. 
 
We now prove that 
$d$ is the eigenvector of $A$, and the eigenvalue is zero. So $Ad=0$,
which means that the one-loop beta function all vanishes, if one assume all the chiral fields have $R$ charge $\frac{2}{3}$.

\textbf{Proof}: Let's first recall some facts about characters \cite{hu2013generalized}. One can define an
inner product on the space of characters, and the characters for irreducible representations form an ortho-normal 
basis: 
\begin{equation*}
  \begin{aligned}
& \sum_C \zeta_C^{-1} \chi_i(C) \chi_j(C^{-1})=\delta_{ij},  \nonumber\\
& \sum_j \chi_j(C^{'}) \chi_j(C^{-1})= \delta_{CC^{'}}\zeta_C\,.
\end{aligned}  
\end{equation*}
Here the first sum is over the conjugacy class, and $\zeta_C$ is the order for the centralizer of one of the elements in $C$. For an arbitrary representation $R$ of finite group $G$, define a matrix $b_{ij}$ by using the tensor product
\begin{equation*}
R\otimes \rho_i=\oplus b_{ij}\rho_j\,.
\end{equation*}
Here $\rho_i$ is the irreducible representation. Now $b_{ij}$ can be represented by the inner product of the representation by using the orthogonal property of the character of the irreducible representation
\begin{equation*}
b_{ij}=\langle R\otimes \rho_i, \rho_j \rangle =\sum_C \zeta_C^{-1} \chi_R(C)\chi_i(C) \chi_j(C^{-1})\,.
\end{equation*}
We compute the action of maxtrix $b$ on vector $\chi(C)=[\chi_1(C), \chi_2(C), \ldots]$, here $C$ is a fixed conjugacy class.
\begin{align*}
&\sum b_{ij} \chi_j(C)= \sum_{j,C^{'}} \zeta_{C^{'}}^{-1} \chi_R(C')\chi_i(C') \chi_j(C^{'-1})\chi_j(C) \nonumber\\
&=\sum_{C^{'}} \chi_R(C')\chi_i(C') \delta_{CC^{'}}=\chi_R(C) \chi_i(C)\,.
\end{align*}
So $\chi(C)$ is the eigenvector of $b$ with eigenvalue $\chi_R(C)$. 
Let's now take $C$ to be the conjugacy class of identity element, and so $\chi(C)=d=[d_1, d_2, \ldots]$ are
the dimension vector, and furthermore $R=3-\rho$ ($\rho$ is the natural representation of $G$ on $\mathbb{C}^3$ ) so that $b_{ij}=B$, and the eigenvalue is just zero!
Similarly, if one take $R=3-\rho^*$ (where $\rho^*$ is the conjugate representation of $\rho$), then the matrix $b_{ij}=(B^T)_{ij}$, and we find the dimension vector 
is also the eigenvector of the matrix $B^T$ whose eigenvalue is zero. 

The above proof also shows that the Mckay quiver does not have non-abelian gauge anomaly, as 
\begin{equation*}
0=(Bd-B^Td)_i=\sum_{j\neq i} a_{ij}d_j-\sum _{j\neq i}a_{ji}d_j\,.
\end{equation*}
Here the first sum counts the number of chirals coming out of the gauge group (fundamental representation), and 
the second term counts the number of chiral coming into the gauge group (anti-fundamental representation). The above formula shows that the number of fundamental chirals are equal to anti-fundamental chirals, and so the gauge anomaly vanishes.

\section{AdS/CFT correspondence}\label{adscftcor}
We will review various physical correspondence appeared in the 4d $\mathcal{N}=1$ AdS/CFT correspondence, 
and emphasize many explicit formulas. The correspondence is summarized in Table \ref{adscft}.
\begin{table}[H]
	\begin{center}
		\begin{tabular}{|c|c|} \hline
			Field theory ${\cal T}$ & Geometry $X$ \\  \hline
                Quiver & Non-commutative crepant resolution (NCCR) \\  \hline
			Quiver Hilbert series $H_{00}(t)$ &  Hilbert series $H(t)$ \\ \hline
                Central charges $a,c$ & Asymptotical expansion of $H(t)$      \\ \hline
			Superconformal index & Kluza-Klein (KK) analysis \\ \hline
                Seiberg duality &  Derived equivalence of NCCR \\ \hline
                Relevant deformation & Modifying geometry\\ \hline
                Exact marginal deformation & Moduli space of Ricci-flat metric on $X$ \\ \hline
		\end{tabular}
	\end{center}
	\caption{AdS/yCFT correspondence for the field theory data and geometry data.}
        \label{adscft}
\end{table}

\subsection{Field theory data}
\label{fieldside}

\textbf{Deformation of $\mathcal{N}=1$ SCFT}:
The bosonic symmetry group of a $\mathcal N = 1$ superconformal algebra is $SO(2, 4) \times U(1)_R \times G_F$. Here $SO(2, 4)$ is the conformal group of four-dimensional Minkowski spacetime, $U(1)_R$ is the R symmetry group, and $G_F$ are other continuous global symmetry groups. A highest weight representation is labeled as $|\Delta, r, j_1, j_2\rangle$, where $\Delta$ is the scaling dimension,
$r$ is $U(1)_R$ charge, $j_1$ and $j_2$ are left and right spin. These states might also carry quantum numbers of flavor symmetry group $G_F$.

Two important short representations are chiral multiplets and multiplets for conserved currents:
\begin{equation}
  \begin{aligned}
      &\mathcal{B}_{r,(j_{1}, 0)}, ~\Delta=\frac{3}{2} r, \\
        &\hat{\mathcal{C}}_{\left(j_{1}, j_{2}\right)}, ~ r=j_{1}-j_{2}, ~ \Delta=2+j_{1}+j_{2}\,.
  \end{aligned}
\end{equation}
Among the multiplets for conserved currents, $\hat{\mathcal{C}}_{(0,0)}$  contains conserved currents for the global symmetry group  $G_{F}$, $\hat{\mathcal{C}}_{\left(\frac{1}{2}, 0\right)}$  contains other supersymmetry currents,  $\hat{\mathcal{C}}_{\left(\frac{1}{2}, \frac{1}{2}\right)}$  contains energy-moment tensor and  $U(1)_{R}$  current, thus $\hat{\mathcal{C}}_{\left(\frac{1}{2}, \frac{1}{2}\right)}$  exists for any  $\mathcal{N}=1$  SCFT.\cite{Xie:2019qmw}

There are only two kinds of supersymmetric deformations for $4d~\mathcal N=1$ SCFT. They can be described as integrating chiral operators over half of the superspace or integrating generic operators over all of the superspace.\cite{green2010exactly} 

For chiral multiplets, we have the following F term relevant or marginal deformation:
\begin{equation}
    \delta S=\int d^2 \theta \lambda \mathcal B_{r,(0,0)}+c.c, \quad r\leq 2\,.
\end{equation}
Here $\lambda$ is a coupling constant. By using conserved current, we also have the following D-term deformations:
\begin{equation}
    \delta S=\int d^2 \theta d^2 \bar\theta \Lambda \hat{\mathcal C}_{(0,0)}\,.
\end{equation}
This is the only relevant or marginal deformation that can be derived from the $ \hat{\mathcal C}$ type operator.

Chiral operators form a ring called the Chiral ring. One can also have continuous moduli space of vacua whose structure is largely determined by the chiral ring \cite{Cachazo:2002ry}.

\textbf{Central charges $a,c$}: The central charges $a$ and $c$ of the SCFT's can be computed from the 't Hooft anomalies $\Tr R$ and $\Tr R^3$ \cite{Anselmi:1997am}:
$$
a=\frac{3}{32}(3 \Tr R^3-\Tr R),\quad c=\frac{1}{32}(9 \Tr R^3-5\Tr R)\,.
$$
For example, the 't Hooft anomalies of a gauge theory with $SU(N)$ gauge group, and coupled with $n_i$ copies of chiral fields in a representation with dimension $r_i$. If the  $U(1)_R$ charge of the matter in $r_i$ representation is $R_i$, then the anomaly is given as \cite{Intriligator:2003jj}:
\begin{equation}
\begin{aligned}
    \Tr R&=(N^2-1)+ \sum_i n_i r_i(R_i-1),\\
    \Tr R^3&=(N^2-1)+ \sum_i  n_i r_i (R_i-1)^3\,.
\end{aligned}
\end{equation}
Here the first term is the contribution from the fermion in vector multiplets, and the second term 
counts the contribution from the chiral multiplets. The $U(1)_R$ charge of the chiral multiplet is 
further constrained by the vanishing of the one-loop beta function, and each term in the superpotential should have 
$U(1)_R$ charge two.

For the McKay quiver described in last section,  the $R$ charges of all chiral multiplets are equal to $\frac{2}{3}$, then the anomalies are
\begin{equation*}
\begin{aligned}
    \Tr R&=\sum_i (d_i^2N^2-1)+\sum_{i\not =j}(\frac{2}{3}-1)a_{ij}d_i d_{j}N^2+\sum_{i}(\frac{2}{3}-1)a_{ii}(d_i^2N^2-1),\\
    \Tr R^3&=\sum_i (d_i^2N^2-1)+\sum_{i\not =j}(\frac{2}{3}-1)^3a_{ij}d_i d_{j}N^2+\sum_{i}(\frac{2}{3}-1)^3a_{ii}(d_i^2N^2-1)\,.
\end{aligned}
\end{equation*}
The $R$ charge condition for the vanishing one-loop beta function is
\begin{equation*}
    6d_i-\sum_j (a_{ij}+a_{ji}) d_j=0\,.
\end{equation*}
Thus
\begin{equation*}
\begin{aligned}
    \Tr R&=\sum_i (d_i^2N^2-1)+\frac{1}{2}\sum_{i\not =j}(\frac{2}{3}-1)(a_{ij}+a_{ji})d_i d_{j}N^2+\sum_{i}(\frac{2}{3}-1)a_{ii}(d_i^2N^2-1)\\
    &=\sum_i (d_i^2N^2-1)+\frac{1}{2}\sum_{i\not =j}(\frac{2}{3}-1)((a_{ij}+a_{ji})d_id_j+2a_{ii}d_i^2)N^2+\sum_{i}(\frac{2}{3}-1)a_{ii}(-1)\\
    &=\sum_i (d_i^2N^2-1)+\frac{1}{2}\sum_{i}(\frac{2}{3}-1)(6d_i)d_i N^2+\sum_{i}(\frac{2}{3}-1)a_{ii}(-1)\\
    &=\sum_i (d_i^2N^2-1)-\sum_{i}d_i^2N^2+\frac{1}{3}L=-Q_0+\frac{1}{3}L,\\
    \Tr R^3&=\sum_i (d_i^2N^2-1)-\sum_{i}\frac{1}{9}d_i^2N^2+\frac{1}{27}L=\frac{8}{9}\sum_i d_i^2 N^2-Q_0+\frac{1}{27}L\,.
\end{aligned}
\end{equation*}
Here $L$ is the number of adjoint multiplets and $Q_0$ is the number of gauge groups.
So the central charges are
\begin{equation*}
\boxed{\begin{aligned}
    a&=\frac{3}{32}(3 \Tr R^3-\Tr R)=\frac{1}{4}\sum_i d_i^2N^2-\frac{1}{48}(L+9Q_0),\\
    c&=\frac{1}{32}(9 \Tr R^3-5\Tr R)=\frac{1}{4}\sum_i d_i^2N^2-\frac{1}{24}(L+3Q_0)\,.
\end{aligned}}
\end{equation*}

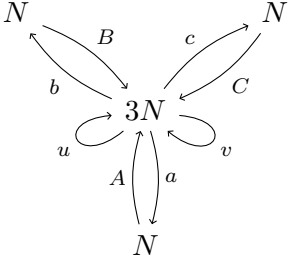
\begin{figure}[H]
	\centering
	\begin{tikzcd}[row sep=normal, column sep=normal]
		N \ar[dr,bend left=15,"B"]&&N\ar[dl, bend left=15, "C"]&\\
		&3N\ar[ul, bend left=15, "b"]\ar[ur, bend left=15, "c"]\ar[d, bend left=15, "a"]\ar[loop,in=185,out=220,looseness=8,"u"]\ar[loop,out=355,in=320,looseness=8,"v"]&& W=uAa+uBb+uCc+u^3+v^3+vAa+vBb+vCc\,.\\[+15pt]
		&N\ar[u, bend left=15, "A"]&&
	\end{tikzcd}  
 \caption{McKay quiver corresponding to Tetrahedral group.}
 \label{Tetrahedral}
\end{figure}

\textbf{Example:} Let's consider the quiver gauge theory shown in Figure \ref{Tetrahedral}. 

\begin{itemize}
\item Vanishing of NSVZ one-loop beta function:
\begin{align*}
& N+\frac{1}{2}(R(A)-1)3N+\frac{1}{2}(R(a)-1)3N=0, \nonumber\\
&N+\frac{1}{2}(R(B)-1)3N+\frac{1}{2}(R(b)-1)3N=0 , \nonumber\\
& N+\frac{1}{2}(R(C)-1)3N+\frac{1}{2}(R(c)-1)3N=0, \nonumber\\
& 3N+3N(R(u)-1)+3N(R(v)-1)+\frac{1}{2}(R(A)-1+R(a)-1)N \nonumber\\
&+\frac{1}{2}(R(B)-1+R(b)-1)N +\frac{1}{2}(R(C)-1+R(c)-1)N=0\,.
\end{align*}
So if all the $R$ charges for the chiral fields are $\frac{2}{3}$, the $R$ charge condition for the 
vanishing one-loop beta function and the superpotential are automatically satisfied.

\item Let's now compute the 't Hooft anomaly by using the above $U(1)_R$ charges:
\begin{equation*}
\begin{aligned}
        &\Tr R=3(N^2-1)+(3N)^2-1+3N^2(\frac{2}{3}-1)\times 6+(9N^2-1)(\frac{2}{3}-1)\times 2=-\frac{10}{3},\\
        &\Tr R^3=3(N^2-1)+((3N)^2-1)+3N^2(\frac{2}{3}-1)^3\times 6+(9N^2-1)(\frac{2}{3}-1)^3\times 2\\
        &=\frac{2(144N^2-53)}{27}\,.
\end{aligned}
\end{equation*}

So the central charges $a,c$ are:
\begin{equation*}
a=\frac{3}{32}(3\Tr R^3-\Tr R)= 3N^2-\frac{19}{24},\quad c=\frac{1}{32}(9\Tr R^3-5\Tr R)= 3N^2-\frac{7}{12} \,.
\end{equation*}
\end{itemize}

\textbf{$a$ maximization and $U(1)_R$ charge}: 
To compute the central charges using the anomaly, one needs to identify the correct $R$ symmetry. Often the IR 
$U(1)_R$ symmetry can not be simply identified with the $U(1)_R$ symmetry of the UV theory, it might mix with 
other abelian global symmetries. One can find out the correct IR $R$ symmetry by using $a$ maximization \cite{Intriligator:2003jj}. The idea is to write the trial $R_{trial}$ as follows:
\begin{equation*}
R_{trial}=R_{UV}+\sum s_IF_I,
\end{equation*}
and then maximize the trial central charge $a_{trial}$ to determine the coefficient $s_I$. 
One must be careful that it is crucial to include all anomaly free symmetries including those 
which can not be seen from the UV theory, i.e. accidental symmetry.

\subsubsection{Superconformal index}
One can define a superconformal index for 4d $\mathcal{N}=1$ theory which can be used to 
extract information about the protected operators \cite{Kinney:2005ej}. Here we summarize the simple large $N$ index 
for the quiver gauge theory proposed in \cite{ginzburg2006calabi, Gadde:2010en}, and 
we will clarify the issue for the choice of $U(N)$ or $SU(N)$ gauge group.

First, we have the single letter indices for $\mathcal{N}=1$ vector multiplet, boson and fermion contribution in a chiral multiplet: 
\begin{equation}
    \begin{aligned}
     i_{V}(t, y) & =\frac{2 t^{6}-t^{3}\left(y+\frac{1}{y}\right)}{\left(1-t^{3} y\right)\left(1-t^{3} y^{-1}\right)}\,,\\
    i_{\phi(r)}(t, y)&=\frac{t^{3r}}{\left(1-t^{3} y\right)\left(1-t^{3} y^{-1}\right)}\,,\\
        i_{\bar{\psi}(r)}(t, y)&=\frac{-t^{3(2-r)}}{\left(1-t^{3} y\right)\left(1-t^{3} y^{-1}\right)}\,,\\
    \end{aligned}
\end{equation}
where $r$ is the R charge for chiral multiplet (the $R$ charge of top component). Then given a quiver gauge theory, we can write down a single letter index matrix $i(t,y)$, whose $(ij)$ entry contains the contribution of the chiral fields 
in bi-fundamental representation $(i\neq j)$, and those chirals in adjoint representation, and contribution of 
vector multiplets:
\begin{align*}
& i(t,y)_{ij}=\sum_{a_{ij}}i_{\phi(r)}(t,y)+\sum_{a_{ji}} i_{\bar{\psi}(r)}(t,y),~\text{when } i\neq j \nonumber\\
& i(t,y)_{ii}=i_V+ \sum_{a_{ii}}i_{\phi(r)}(t,y)+\sum_{a_{ii}} i_{\bar{\psi}(r)}(t,y)\,.
\end{align*}

One can find  the single trace index from matrix $i(t,y)$ as follows \cite{Gadde:2010en} (assuming the gauge group 
is $U(N)$):
\begin{equation}
    \begin{aligned}
     \mathcal I_{\text{s.t.}}=-\sum_{k=1}^\infty \frac{\varphi(k)}{k}\log[\det(1-i(t^k,y^k))],
    \end{aligned}
\end{equation}
where $\varphi(n)$ is the Euler Phi function and we can use the following property to calculate it
\begin{equation}
    \begin{aligned}
     \sum_{k=1}^\infty \frac{\varphi(k)}{k}\log(1-x^k)=\frac{-x}{1-x}\,.
    \end{aligned}
\end{equation}

The matrix $i(t,y)$ could be simplified as follows \cite{eager2012superconformal}. Notice that 
\begin{equation*}
1-i_V(t,y)={1-t^6 \over \left(1-t^{3} y\right)\left(1-t^{3} y^{-1}\right)},
\end{equation*}
so the matrix $1-i(t,y)$ can be written as
\begin{equation*}
1-i(t,y)={\chi(t) \over \left(1-t^{3} y\right)\left(1-t^{3} y^{-1}\right)}
\end{equation*}
with $\chi(t)$ the matrix 
\begin{equation}
\boxed{\chi(t)=1-M_Q(t)+t^6M_Q^T(t^{-1})-t^6},
\label{universal}
\end{equation}
and $M_Q(t)$ is the matrix counting the $R$ charge of the chiral fields:
\begin{enumerate}
\item If $i\neq j$, the matrix element $M_{ij}$ of $M_Q$ is  $$M_{ij}=\sum_{\text{bifundamental chiral fields in} (\mathbb{N}_i,\mathbb{\bar{N}}_j)}t^{3R_{ij}}\,.$$
\item If $i=j$, the matrix element $M_{ii}$ is 
$$
M_{ii}=\sum_{\text{adjoint chiral fields}}t^{3R_{ii}}\,.
$$
\end{enumerate}
The matrix $\chi(t)$ was also discovered by Ginzburg in \cite{ginzburg2006calabi}. The single trace is simplified as 
\begin{equation}
\mathcal I_{\text{s.t.}}=-\sum_{k=1}^\infty \frac{\varphi(k)}{k}\log[\det( \chi(t^k))]+Q_0 i_V\,.
\label{final}
\end{equation}
Here $Q_0$ is the number of quiver nodes.  Next, we would like to take the gauge group as $SU$ type instead of 
$U$ type, and we need to subtract the contributions of the free fields in the IR. Firstly, there is a $U(1)$ vector multiplet for each gauge group, so the second term in (\ref{final}) is subtracted; Secondly, there might be 
free chirals (For example, if there are adjoint chirals, then the traceless part of it is free and decoupled); Thirdly, there could also be some other free chirals which one need to subtract (sometimes one can find them from the explicit $U(1)_R$ charge in the quiver gauge theory), and the best way to find them is to expand the first term in \ref{final} and detect the number of free chirals. So the final single trace index takes the form
\begin{equation}
\boxed{\mathcal I_{\text{s.t.}}=-\sum_{k=1}^\infty \frac{\varphi(k)}{k}\log[\det( \chi(t^k))]-\text{free chirals}}\,.
\end{equation}
An even subtle situation is that one finds from the $U$ type index the contribution of fields violating the unitarity bound, then one needs to be more careful. 

\textbf{Example}: Let's consider the quiver gauge theory in Figure \ref{Tetrahedral}. The matrix $M_Q$ is
$$M_Q=\begin{pmatrix}
    0&0&t^{2}&0\\
    0&0& t^{2}&0\\
    t^{2}&t^{2}&2t^{2}&t^{2}\\
     0&0& t^{2}&0
\end{pmatrix}\,.$$
Here the $R$ charges of the chiral fields are all $\frac{2}{3}$, and the matrix $\chi(t)$ is 
$$\chi(t)=\begin{pmatrix}
    1-t^6&0&t^4-t^{2}&0\\
    0&1-t^6& t^4-t^{2}&0\\
    t^4-t^{2}&t^4-t^{2}&1-t^6+2(t^4-t^{2})&t^4-t^{2}\\
     0&0& t^4-t^{2}&1-t^6
\end{pmatrix}\,.$$

Then we can calculate the single trace index (for $U$ type gauge group) (see \ref{final})
\begin{equation}
    \begin{aligned}
    &\det(\chi(t))=(1-t^2)^2(1-t^4)^2(1-t^6)^2\,,\\
    &\mathcal I_{\text{s.t.}}=-\sum_{k=1}^\infty \frac{\varphi(k)}{k}\log[(1-t^{2k})^2(1-t^{4k})^2(1-t^{6k})^2]+4i_V\\
    &=2\frac{t^2}{1-t^2}+2\frac{t^4}{1-t^4}+2\frac{t^6}{1-t^6}+4i_V\,.
    \end{aligned}
\end{equation}
To get the index for $SU$ type gauge group, we notice that there are two free chirals from the two adjoints, so the index for $SU$ type theory is 
\begin{equation*}
\mathcal I_{\text{s.t.}}=2\frac{t^2}{1-t^2}+2\frac{t^4}{1-t^4}+2\frac{t^6}{1-t^6}-2\frac{t^2-t^4}{(1-t^3y)(1-t^3y^{-1})}\,.
\end{equation*}

The full index can be computed by calculating plethystic exponential of the single trace index, i.e.
\begin{equation*}
\mathcal I(t,y)=PE(\mathcal I_{\text{s.t.}}(t,y))=\exp^{\sum_n \frac{1}{n}\mathcal I_{\text{s.t.}}(t^n,y^n)},
\end{equation*}
and the first fewer terms in the expansion could tell us the information of relevant and exact marginal deformations. For example, the expansion of the full index of the above theory is  
\begin{equation*}
\mathcal I_{\text{full}}=1+6 t^4-2 t^5 \left(y+\frac{1}{y}\right)+4 t^6+O\left(t^7\right)\,.
\end{equation*}
So there should be 6 chiral scalar operators with $U(1)_R$ charge $\frac{4}{3}$ (those can be identified as $\Tr(Aa), \Tr(Bb)$ etc, and 4  marginal operators.
Notice that the conserved current for the global symmetry 
and the marginal chiral scalar operator give opposite contributions to the index \cite{Beem:2012yn,green2010exactly}, so one can get the information of the exact marginal operators (one needs to include the flavor fugacity to the full index to determine exactly the number of exact marginal operators).

\subsubsection{Quiver Hilbert series and dual geometry}
Given a quiver gauge theory, we can calculate a matrix Hilbert series \cite{bocklandt2008graded} as follows: 
$$
H(Q,t)=\frac{1}{1-M^\prime_Q(t)+t^2{M^\prime_Q}^T(t^{-1})-t^2}\,.
$$
Here adjacent matrix $M'_Q$ can be read from the quiver and R-charges:

\begin{enumerate}
\item If $i\neq j$, the matrix element $M'_{ij}$ of $M'_Q$ is  $$M'_{ij}=\sum_{\text{bifundamental chiral fields in} (\mathbb{N}_i,\mathbb{\bar{N}}_j)}t^{R_{ij}}\,.$$
\item If $i=j$, the matrix element $M'_{ii}$ is 
$$
M'_{ii}=\sum_{\text{adjoint chiral fields}}t^{R_{ii}}\,.
$$
\end{enumerate}
Here $R_{ij}$ is the $R$ charge of a bifundamental chiral field in $(\mathbb{N}_i,\mathbb{\bar{N}}_j)$. The meaning of the entry $H_{ij}$ counts the oriented path from the node $i$ to node $j$ with the $R$ charge grading.
The Hilbert series $H_{00}(Q,t)$ is the 00-component of the matrix $H(Q,t)$, which counts the gauge invariant scalar operators (closed loop in the quiver which passes through node 0). In our case, there is always a node corresponding to the trivial representation, which we take as the node $0$.

The $H_{00}$ is conjectured to be identified with the Hilbert series of the dual geometry, from which one 
may guess the dual geometry. 

\textbf{Example}: Let's look at the quiver shown in Figure \ref{Tetrahedral}, the  matrix $M'_Q$ is 

$$M'_Q=\begin{pmatrix}
    0&0&t^{2/3}&0\\
    0&0& t^{2/3}&0\\
    t^{2/3}&t^{2/3}&2t^{2/3}&t^{2/3}\\
     0&0& t^{2/3}&0
\end{pmatrix}$$
and so the matrix Hilbert series is
$$
H_{00}=\frac{1-t^{8}}{(1-t^2)(1-t^{8/3})(1-t^4)(1-t^{4/3})}\,.
$$

$H_{00}$ takes the same form as that of a three dimensional hypersurface singularity, and the weights read
$(d;x,y,z,w)=(1;\frac{1}{2},\frac{1}{3},\frac{1}{4},\frac{1}{6})$. The geometry might take the following form:
$$
x^2+y^3+z^4+w^2y^2+w^2z^2+wyz^2=0\,.
$$
Of course, one cannot completely determine the singularity just from the Hibert series. More information is needed to find out the dual geometry.

The quiver Hilbert series $H_{00}$ counts  chiral scalar operators formed by loops passing through node $0$. For example, the first fewer terms in the $t$ expansion of $H_{00}$ would tell us some information about the relevant and  marginal deformations:
\begin{equation*}
H_{00}(t)=1+t^{4/3}+t^2+2 t^{8/3}+O\left(t^{10/3}\right)\,.
\end{equation*}
This tells us that the theory has a relevant operator with $R$ charge $\frac{4}{3}$, and 
a marginal operator with $R$ charge $2$. 

\subsubsection{Seiberg duality}
Seiberg duality identifies the IR SCFT with two different UV non-abelian gauge theories. The original example involves $SU(N_c)$ gauge theory coupled with $N_f$ fundamental matter $(Q, \tilde{Q})$. The dual theory is 
$SU(N_f-N_c)$ coupled with $N_f$ fundamental matter, and additionally there is uncharged chiral multiplet $M$ coupled through a cubic superpotential.

The basic idea of Seiberg duality can be generalized to quiver gauge theories. The rules of Seiberg duality for the quiver are summarized as follows \cite{Seiberg:1994pq,derksen2008quivers,nolla2017flops}. Here we consider the Seiberg duality acting on a node $k$ with gauge group $SU(N)$: 
\begin{enumerate}
 \item  Add a new arrow $[ab]$ from node $i$ and node $j$ for each oriented path $i\to k \to j$, and the corresponding chiral fields are labeled as $
    \begin{tikzcd}
	i & k & j
	\arrow["a", from=1-1, to=1-2]
	\arrow["b", from=1-2, to=1-3]
\end{tikzcd}
    $.

\item Change the rank of gauge group as $N\rightarrow F-N$. Here $F$ is the number of bi-fundamental fields on node $k$ divided by two.

\item Change the direction of all the arrows attached to this node $k$: $a\rightarrow a^*,b\rightarrow b^*$.

\item The superpotential is changed as $W^{\prime}=[W]+\Delta$: $[W]$ is defined by replacing all fields combination $ab$  appearing in $W$ by the new field $[ab]$, and
$\Delta=\sum [ab]b^*a^*$.

\item Do the reduction by integrating out massive fields.

\end{enumerate}

Let's give a simple example first. The original quiver is shown in Figure \ref{exsd}. The superpotential is $W=abc$. 
\begin{figure}[H]
    \centering
    \begin{tikzcd}
	F1 & N & F2
	\arrow["a", from=1-1, to=1-2]
	\arrow["b", from=1-2, to=1-3]
	\arrow["c", curve={height=-12pt}, from=1-3, to=1-1]
\end{tikzcd}
    \caption{A quiver with superpotential $W=abc$. Seiberg duality is done on the node with rank $N$.}
    \label{exsd}
\end{figure}
Now we do the Seiberg duality on node $N$. We should replace $a,b$ by $a^*,b^*$,  and add a new field $[ab]$. The rank of gauge group is changed to $\frac{F1+F2}{2}-N$. The resulting quiver diagram is shown in Figure \ref{exsd1}.

\begin{figure}[H]
    \centering
    \begin{tikzcd}
	{F1} & {\frac{F_1+F_2}{2}-N} & {F2}
	\arrow["{a^*}", from=1-2, to=1-1]
	\arrow["{b^*}", from=1-3, to=1-2]
	\arrow["c"', curve={height=-18pt}, from=1-3, to=1-1]
	\arrow["{[ab]}"{description}, curve={height=-18pt}, from=1-1, to=1-3]
\end{tikzcd}
\caption{The dual quiver after doing Seiberg duality on the central node of Figure \ref{exsd}.}
    \label{exsd1}
\end{figure}

The new superpotential is
\begin{equation*}
    W=abc\rightarrow W^{\prime}=[W]+\Delta=[ab]c+[ab]b^*a^*\,.
\end{equation*}

Finally, we do the reduction by solving the fields $[ab]$ and $c$ in terms of other fields
\begin{align*}
    &\frac{\partial W'}{\partial[ab]}=c+b^*a^*=c+b^*a^*=0, \nonumber\\
&\frac{\partial W'}{\partial c}=[ab]=0\,.
\end{align*}
The final quiver is shown in Figure \ref{exsd2} and the dual superpotential is $W=0$.
\begin{figure}[H]
    \centering
    \begin{tikzcd}
	F1 & {\frac{F_1+F_2}{2}-N} & F2
	\arrow["{b^*}", from=1-3, to=1-2]
	\arrow["{a^*}", from=1-2, to=1-1]
\end{tikzcd}
\caption{The final quiver after the reduction on the quiver depicted in Figure \ref{exsd1}.}
    \label{exsd2}
\end{figure}

\begin{figure}[H]
	\centering
	\begin{tikzcd}[row sep=normal, column sep=normal]
		* \ar[dr,bend left=15,"B"]&&N\ar[dl, bend left=15, "C"]\\
		&3N\ar[ul, bend left=15, "b"]\ar[ur, bend left=15, "c"]\ar[d, bend left=15, "a"]\ar[loop,in=185,out=220,looseness=8,"u"]\ar[loop,out=355,in=320,looseness=8,"v"]&\\[+15pt]
		&N\ar[u, bend left=15, "A"]&\\[-15pt]
		&\textbf{Q}&
	\end{tikzcd}  
	\begin{tikzcd}
		\stackrel{}{\Longrightarrow}    
	\end{tikzcd}
	\begin{tikzcd}[row sep=normal, column sep=normal]
		* \ar[dr,bend left=15,"b^*"]&&N\ar[dl, bend left=15, "C"]\\
		&3N\ar[ul, bend left=15, "B^*"]\ar[ur, bend left=15, "c"]\ar[d, bend left=15, "a"]\ar[loop,in=185,out=220,looseness=8,"u"]\ar[loop,out=355,in=320,looseness=8,"v"]\ar[loop,in=115,out=78,looseness=8,"{[Bb]}"]&\\[+15pt]
		&N\ar[u, bend left=15, "A"]&\\[-15pt]
		&\textbf{ }&
	\end{tikzcd}  
	\begin{tikzcd}
		\stackrel{}{\Longrightarrow}    
	\end{tikzcd}
	\begin{tikzcd}[row sep=normal, column sep=normal]
		2N \ar[dr,bend left=15,"b"]&&N\ar[dl, bend left=15, "C"]\\
		&3N\ar[ul, bend left=15, "B"]\ar[ur, bend left=15, "c"]\ar[d, bend left=15, "a"]\ar[loop,in=185,out=220,looseness=8,"u"]&\\[+15pt]
		&N\ar[u, bend left=15, "A"]&\\[-15pt]
		&\textbf{Q}_1&
	\end{tikzcd}  
	\caption{Seiberg duality of the quiver of the Tetrahedral group. Here the duality is done on the star node.}
        \label{extetrahedral}
\end{figure}
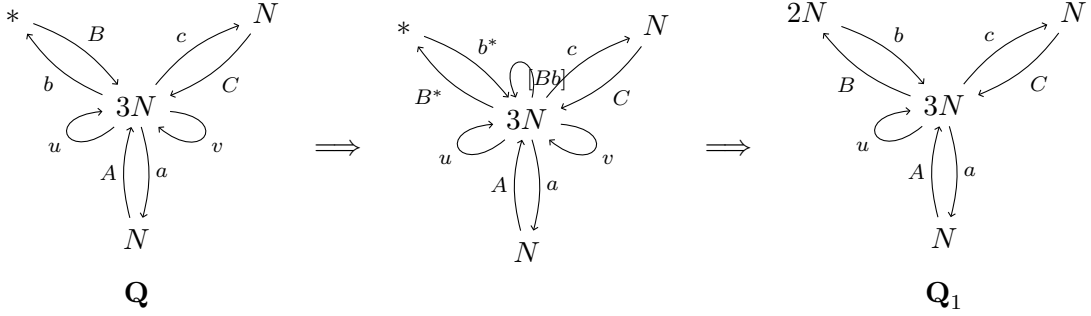

For a more complicated example, we take a look at quiver gauge theory \textbf{Q} in Figure \ref{extetrahedral}.  We consider the Seiberg duality acting on the node $*$.
The superpotential is
\begin{equation}
    \begin{aligned}
        W=uAa+\omega uBb+\omega^2 uCc-\frac{1}{3}u^3-vAa-\omega^2vBb-\omega vCc+\frac{1}{3}v^3,
    \end{aligned}
\end{equation}
where  $\omega=e^{2\pi i/3}$ is a constant.

According to the procedure outlined above, we first add a new adjoint field $[Bb]$ on the middle node, and change the rank of $*$ to $2N$. Then change the direction of $B$ and $b$, and denote them as $B^*,b^*$ respectively. Now to get the new superpotential $W'$, we replace the term $Bb$ in $W$ by $[Bb]$ and add $[Bb]B^*b^*$ term:
$$
W'=[W]+\Delta=uAa+\omega u[Bb]+\omega^2 uCc-\frac{1}{3}u^3-vAa-\omega^2v[Bb]-\omega vCc+\frac{1}{3}v^3
+[Bb]B^*b^*\,.
$$
Next we do the reduction by integrating out massive superfields. By doing partial differentiation, we can get linear terms of $[Bb],v$:
\begin{equation}
    \begin{aligned}
\frac{\partial W'}{\partial [Bb]}&=\omega u-\omega^2 v+B^*b^*=0,\\
\frac{\partial W'}{\partial v}&=-Aa-\omega^2[Bb]-\omega Cc+v^2=0\,.
    \end{aligned}
\end{equation}
Then we can replace $[Bb]$ and $v$ by other terms
\begin{equation}
    \begin{aligned}
     v&=\omega^2 u+\omega B^*b^*\,,\\
     [Bb]&=-\omega Aa-\omega^2Cc+\omega v^2\,,
    \end{aligned}
\end{equation}
and rename $B^*,b^*$ by $B,b$. Thus the final result is
$$
W^{\prime}\to W^{\prime}=(1-\omega^2)Aau-wAaBb+(\omega^2-1)Ccu-\omega^2 CcbB+\omega^2Bbu^2+\omega (Bb)^2 u+\frac{1}{3}(Bb)^3\,.
$$
The whole process is shown in Figure \ref{extetrahedral}.

\subsection{Gravity side}\label{gravityside}
Let's now turn to the gravity side where much information is contained in the geometry of the singularity $X$. 
We constrain $X$ to be a Gorenstein canonical singularity with a $C^*$ action, and furthermore it is required to be stable \cite{collins2019sasaki,collins2016k}.
Since the theory is determined by the local behavior of the singularity, one just needs to know the affine ring 
attached to the singularity. 

\textbf{$U(1)_R$ symmetry and $\mathbb{C}^*$ action}:
4d $\mathcal{N}=1$ SCFT has a $U(1)_R$ symmetry, which is identified with the $C^*$ action on $X$, therefore one 
gets a graded affine ring from the dual geometry. Since the embedding coordinates would give rise to chiral operators
of the field theory, the weights of the $C^*$ action on the embedding coordinates should be positive. 

To identify the normalization constant, one needs to impose the generators for the volume form $\Omega$ to have $R$
charge two
\begin{equation*}
[\Omega]=2\,.
\end{equation*}
Such homogeneous generator $\Omega$ exists as $X$ is assumed to be Gorenstein.

\textbf{Hilbert series and central charges}:
Given a graded ring $R$, one can define a Hilbert series as follows 
\begin{equation*}
H(t)=\sum \text{dim} H_\alpha t^\alpha\,.
\end{equation*}
Here $\text{dim} (H_\alpha)$ denotes the dimension of the elements with $C^*$ charge $\alpha$. If $R$ is a complete intersection, the Hilbert series takes a simple form:
$$
H(t)=\frac{(1-t^{d_1})\cdots(1-t^{d_n})}{(1-t^{w_1})\cdots(1-t^{w_s})}
$$
where $d_i,i=1,\cdots,n$ are degrees of defining ideal $f_i$, and $w_j,j=1,\cdots,s$ are weights of the coordinates $x_j$.

For example,  let's take $X$ to be given by the hypersurface singularity $x^2+y^2+z^4+w^4=0$. Then the coordinate ring of $X$ is $R_X=\frac{\mathbb{C}[x,y,z,w]}{\{x^2+y^2+z^4+w^4=0\}}$. The
weights of $x,y,z,w$ are $(x,y,z,w)=(2,2,1,1)$, and the weight of the hypersurface is $4$.
The Hilbert series is $H(t)=\frac{1-t^4}{(1-t)^2(1-t^2)^2}$.
The operators with different degrees in the graded ring X are listed as follows:
$$\begin{array}{ll}
    H_{1}:z,w & \text{dim} H_1=2 \\
    H_{2}:z^2,w^2,zw,x,y & \text{dim} H_2=5\\
    H_3: z^3,w^3,zw^2,z^2w,xz,xw,yz,yw & \text{dim}H_3 =8
\end{array}$$
Then we can expand the Hilbert series and check the first few terms:
\begin{align*}
&H(t)=\frac{1-t^4}{(1-t)^2(1-t^2)^2}=1+2t+5t^2+8t^{3}+\cdots \nonumber\\
&=1+\text{dim} H_{1} t+\text{dim} H_{2} t^{2}+ \text{dim} H_3 t^3+\cdots=\sum_{\alpha}\text{dim} H_{\alpha}t^{\alpha}\,.
\end{align*}

The central charge $a,c$ can be extracted from the Hilbert series as follows \cite{Martelli:2005tp}. Let's define $t=\exp(-s)$, then 
$H(t)$ has the expansion
\begin{equation*}
H(\exp(-s))={a_0\over s^3}+\frac{a_1}{s^2}+\ldots\,.
\end{equation*}
Then the central charge of the dual theory is given as
\begin{equation*}
a=c=\frac{27}{32}\frac{1}{a_0}N^2\,.
\end{equation*}
Notice that this central charge is just the value in the leading order of $N$. When there are more than one 
candidate $U(1)_R$ symmetry, one needs to minimize the trial central charge $a_0(R_i)$ to find the correct $U(1)_R$ symmetry \cite{Martelli:2005tp}, where $R_i$ is some unfixed $R$ charge. 

When $X$ defines an isolated singularity whose link $L$ gives the internal geometry of the dual SCFT, the elements in the coordinate ring of $X$ would give the chiral scalar operators, see \cite{Xie:2019qmw}.

\textbf{Example}: Consider a quasi-homogeneous hypersurface $f$ whose weights are $(w_1, w_2, w_3, w_4; d)$, and the top form is 
\begin{equation*}
\Omega= {dx \wedge dy \wedge dz \wedge dw \over df}\,.
\end{equation*}
To require the weight of $\Omega$ to be two, the normalization constant must be 
\begin{equation}\label{normalization}
(w_1+w_2+w_3+w_4-d) \delta =2~~\to~\delta= \frac{2}{w_1+w_2+w_3+w_4-d}\,.
\end{equation}
By computing the Hilbert series and doing the expansion, we find the coefficient $a_0$ is 
\begin{equation*}
a_0=\frac{d}{w_1 w_2 w_3 w_4}\delta^{-3}\,.
\end{equation*}
So the central charge is given as 
\begin{equation}\label{cc}
a=c={27\over 32}{w_1 w_2 w_3 w_4\over d}\delta^3 N^2\,.
\end{equation}

\textbf{Exact marginal deformations}: One can find exact marginal deformations from the geometry side as follows: a): there is always a type IIB axio-dilaton field $\tau$; b) There are ones coming from the $H^2(L)$ of the link $L$, which 
can be computed using the data of $X$; c): If there is a moduli space for the Ricci-flat metric, then one 
gets exact marginal deformations from those moduli; If $X$ is defined as the hypersurface $f$, one can easily find out those deformations by counting deformations with the same weight as $f$.

\textbf{Baryonic symmetry}: There are other global symmetries besides the $U(1)_R$ symmetry. On the gravity side, this comes from the massless gauge field on the bulk. A simple source of such symmetry is also a nonzero value of $H^2(L)$.

\textbf{Index and KK analysis}: To get information of $\frac{1}{4}$ BPS operators, one needs to perform the detailed Kluza-Klein (KK) analysis, see \cite{ceresole1999kk,ceresole2000spectrum,eager2012superconformal} for the case when $L$ is smooth. In our case, most of the dual geometry $S_5/G$ have singularities so extra care is needed.
The details will appear elsewhere.

\subsubsection{Smoothing of the singularity}
\textbf{Deformation and new AdS/CFT pair}: One can smooth the singularity $X$ ( making it less singular) by changing 
the defining equation. In some cases, one might find a new geometry $X^{'}$ which also has a nice dual SCFT.
Sometimes the change of the geometry can be described by the relevant deformation of the original field theory.
Typical example is the affine $A_1$ $\mathcal{N}=2$ theory, whose dual geometry is $x^2+y^2+z^2=0, w$;
after turning on the mass deformation, one gets a conifold quiver,  and the dual geometry $X^{'}$ is given 
by $x^2+y^2+z^2+w^2=0$ which is regarded as the deformation of $X$. We will discuss more examples in this paper.

\textbf{Quiver and crepant Resolution }
One can smooth the singularity by doing the resolution of singularity. 
To preserve supersymmetry, one needs to study a special kind of resolution called crepant resolution \cite{reid46young}. The end of crepant resolution might still have singularities (so that there is only partial resolution). The quiver gauge theory is related to a closed related resolution called non-commutative crepant resolution (NCCR)\cite{van2004non}. In fact, it seems that the physical quiver gauge theory should be regarded as the mathematically studied quiver.

\section{Finite subgroups of $SU(2)$}
\label{finitesubgpsu2}

The finite subgroups of the special unitary group $SU(2)$ have an $ADE$ classification. For each subgroup $G$, we list the equation corresponding to $\mathbb{C}^2/G \times \mathbb{C}$ and the physical data such as leading order of central charge $a,c$ and Hilbert series $\frac{1-t^d}{(1-t^{w_1})(1-t^{w_2})(1-t^{w_3})(1-t^{w_4})}$ from the geometry side in Table \ref{su2subgroup}. We draw the quiver diagrams in Table \ref{su2quiver} and list the corresponding field theory data in Table \ref{quivergaugeSu2}. The corresponding quiver gauge theory was well-known \cite{douglas1996d}, and here the index and the quiver Hilbert series are 
computed. The $R$ charges cannot be fixed by equation (\ref{normalization}) solely, which is different from the example we have discussed in section \ref{gravityside} and Table \ref{su2subgroupwithoutloops}. The $R$ charge of $w$ can be fixed by minimizing the $a_0(R_w)$ as we mentioned in section \ref{gravityside}, and the result is $R_w=\frac{2}{3}$.

Given the branching rules of the irreducible representations of  $ADE$ groups, one can figure out the quiver diagrams as shown in Table \ref{su2quiver} by McKay correspondence as we introduced in section \ref{mkproperty}. Field theory data such as central charges $a,c$, quiver Hilbert series, and single trace index can be computed using the $R$ charges. The calculation method is introduced in section \ref{fieldside} and the results are summarized in Table \ref{quivergaugeSu2}.

\begin{table}[H]
\begin{center}
\resizebox{1\columnwidth}{!}{
\begin{tabular}{|c|c|l|c|c|c|c|} \hline 
Group $G$ & Order $|G|$&Name &  Equation  &$R$ charge $(d;w_x,w_y,w_z,w_w)$ & Hilbert series $H_{00}$& Leading order of $a,c$\\ \hline 
Cyclic $\mathbb{Z}_{n+1}$ &n+1&   $A_{n}$ & $x^2+y^2+z^{n+1}=0$  &$\left\{\frac{4 (n+1)}{3};\frac{2 (n+1)}{3},\frac{2 (n+1)}{3},\frac{4}{3},\frac{2}{3}\right\} $&$H_{00}=\frac{1-t^{4(n+1)/3)}}{(1-t^{2(n+1)/3})(1-t^{2(n+1)/3})(1-t^{4/3})(1-t^{2/3})}$ & $\frac{n+1}{4}N^2$\\ \hline  
Binary dihedral $\mathbb{D}_{n}$ & 4(n-2)&  $D_{n}$& $x^2+y^{n-1}+yz^2=0$  & $\left\{\frac{8 (n-1)}{3};\frac{4 (n-1)}{3},\frac{8}{3},\frac{4 (n-2)}{3},\frac{2}{3}\right\}$ & $H_{00}=\frac{1-t^{8(n-1)/3}}{(1-t^{4(n-1)/3})(1-t^{8/3})(1-t^{4(n-2)/3})(1-t^{2/3})}$&$(n-2)N^2$ \\   \hline 
 Binary tetrahedral $\mathbb{T}$ &24& $E_6$  & $x^2+y^3+z^4=0$  &  $\left\{16;8,\frac{16}{3},4,\frac{2}{3}\right\}$ & $H_{00}=\frac{1-t^{16}}{(1-t^{8})(1-t^{16/3})(1-t^4)(1-t^{2/3})}$&$6N^2$  \\ \hline 
  Binary octahedral  $\mathbb{O}$ & 48&$E_7$  & $x^2+y^3+yz^3=0$ & $\left\{24;12,8,\frac{16}{3},\frac{2}{3}\right\}$ &$H_{00}=\frac{1-t^{24}}{(1-t^{12})(1-t^{8})(1-t^{16/3})(1-t^{2/3})}$  &$12 N^2$ \\ \hline 
 Binary icosahedral $\mathbb{I}$ & 120&$E_8$  &$x^2+y^3+z^5=0$  &   $\left\{40;20,\frac{40}{3},8,\frac{2}{3}\right\}$  & $H_{00}=\frac{1-t^{40}}{(1-t^{20})(1-t^{40/3})(1-t^8)(1-t^{2/3})}$& $30N^2$ \\ \hline 
\end{tabular}}
\end{center}
\caption{Geometric data of the corresponding singularity $\mathbb{C}^2/G( G=ADE)$ whose dual quiver is the affine ADE quiver, and the R charge is found by requiring the top form to have charge $2$ and $a_0(R_w)$ minimization.}
\label{su2subgroup}
\end{table}


\begin{table}[H]
  \begin{center}
	\scalebox{0.6}{
    \begin{tabular}{|c|c|}
\hline
Group& Mckay quiver\\
    \hline
   $A_n$&     
\begin{tikzcd}[ampersand replacement=\&] 
	\&N\ar[dl,bend left=10,""]\ar[dr,bend left=10,""]\ar[loop,out=105,in=75,looseness=9,""]\&\\[-5pt]
	N\ar[ur,bend left=10,""]\ar[d,bend left=10,""]\ar[loop,out=165,in=135,looseness=9,""]\&\&N\ar[ul,bend left=10,""]\ar[d,bend left=10,""]\ar[loop,out=45,in=15,looseness=9,""]\\[+10pt]
	N\ar[dr,bend left=10,""]\ar[u,bend left=10,""]\ar[loop,out=225,in=195,looseness=9,""]\&\&N\ar[dl,bend left=10,dashed, no head]\ar[u,bend left=10,""]\ar[loop,out=345,in=315,looseness=9,""]\\[-5pt]
	\&N\ar[ul,bend left=10,""]\ar[loop,out=285,in=255,looseness=12,""]\&\\
\end{tikzcd}  \\
\hline
$D_n$&\begin{tikzcd}[ampersand replacement=\&]
	N\ar[dr,bend left=10,""]\ar[loop,out=105,in=75,looseness=9,""]\&\&\&\&\&N\ar[dl,bend left=10,"" ]\ar[loop,out=105,in=75,looseness=9,""]\\
	\&2N\ar[ul,bend left=10,""]\ar[dl,bend left=10,""]\ar[r,bend left=10,""]\ar[loop,out=105,in=75,looseness=9,""]
	\&2N\ar[l,bend left=10,""]\ar[r,dashed,no head]\ar[loop,out=105,in=75,looseness=9,""]
	\&2N\ar[r,bend left=10,""]\ar[loop,out=105,in=75,looseness=9,""]
	\&2N\ar[l,bend left=10,""]\ar[ur,bend left=10,""]\ar[dr,bend left=10,""]\ar[loop,out=105,in=75,looseness=9,""]\\
	N\ar[ur,bend left=10,""]\ar[loop,out=105,in=75,looseness=9,""]\&\&\&\&\&N\ar[ul,bend left=10,""]\ar[loop,out=105,in=75,looseness=9,""]\\
\end{tikzcd} \\
\hline

$E_6$&

\begin{tikzcd}[ampersand replacement=\&]
	\&\&N\ar[d,bend left=10,""] \ar[loop,out=105,in=75,looseness=9,""]\\
	\&\&2N\ar[u,bend left=10,""]\ar[d,bend left=10,""]\ar[loop,out=10,in=350,looseness=9,""]\\ N\ar[r,bend left=10,""]\ar[loop,out=105,in=75,looseness=9,""]\&2N\ar[r,bend left=10,""]\ar[l,bend left=10,""]\ar[loop,out=105,in=75,looseness=9,""]\&3N\ar[u,bend left=10,""]\ar[r,bend left=10,""]\ar[l,bend left=10,""]\ar[loop,out=285,in=250,looseness=9,""]\&2N\ar[r,bend left=10,""]\ar[l,bend left=10,""]\ar[loop,out=105,in=75,looseness=9,""]\&N\ar[l,bend left=10,""]\ar[loop,out=105,in=75,looseness=9,""]
\end{tikzcd}\\
\hline   

$E_7$&

\begin{tikzcd}[ampersand replacement=\&]
	\&\&\&2N\ar[d,bend left=10,""] \ar[loop,out=105,in=75,looseness=9,""]\\
 N\ar[r,bend left=10,""]\ar[loop,out=105,in=70,looseness=9,""]\&2N\ar[r,bend left=10,""]\ar[l,bend left=10,""] \ar[loop,out=105,in=75,looseness=9,""]\&3N\ar[r,bend left=10,""]\ar[l,bend left=10,""] \ar[loop,out=105,in=75,looseness=9,""]\&4N\ar[r,bend left=10,""]\ar[l,bend left=10,""]\ar[u,bend left=10,""] \ar[loop,out=285,in=250,looseness=9,""] \&3N\ar[r,bend left =10,""]\ar[l,bend left=10,""] \ar[loop,out=105,in=75,looseness=9,""]\&2N\ar[r,bend left=10,""]\ar[l,bend left=10,""]\ar[loop,out=105,in=75,looseness=9,""] \&N\ar[l,bend left=10,""]\ar[loop,out=105,in=75,looseness=9,""]
\end{tikzcd}\\
\hline
  $E_8$&

\begin{tikzcd}[ampersand replacement=\&]
	\&\&\&\&\&3N\ar[d,bend left=10,""] \ar[loop,out=105,in=75,looseness=9,""]\\
 N\ar[r,bend left=10,""]\ar[loop,out=105,in=75,looseness=9,""]\&2N\ar[r,bend left=10,""]\ar[l,bend left=10,""] \ar[loop,out=105,in=75,looseness=9,""]\&3N\ar[r,bend left=10,""]\ar[l,bend left=10,""] \ar[loop,out=105,in=75,looseness=9,""]\&4N\ar[r,bend left=10,""]\ar[l,bend left=10,""]\ar[loop,out=105,in=75,looseness=9,""] \&5N\ar[r,bend left =10,""]\ar[l,bend left=10,""] \ar[loop,out=105,in=75,looseness=9,""]\&6N\ar[u,bend left=10,""]\ar[r,bend left=10,""]\ar[l,bend left=10,""]\ar[loop,out=285,in=250,looseness=9,""] \&4N\ar[l,bend left=10,""]\ar[r,bend left=10,""]\ar[loop,out=105,in=75,looseness=9,""]
\&2N\ar[l,bend left=10,""]\ar[loop,out=105,in=75,looseness=9,""]
\end{tikzcd}  \\
\hline 
    \end{tabular}}
   \end{center}
    \caption{McKay quivers corresponding to finite subgroups of $SU(2)$ group.}
    \label{su2quiver}
\end{table}

\begin{table}[htbp]
	\begin{center}
		\renewcommand\arraystretch{2.2}
		\scalebox{0.8}{
			\begin{tabular}{|l|l|l|} \hline
				Group & Field theory data&\\  \hline
			\multirow{3}{*}{\makecell{ $A_n$}} &
				Central charges $(a,c)$ & $a=\frac{1}{24}(1+n)(-5+6N^2)\quad c=\frac{1}{12}(1+n)(-2+3N^2)$
				\\ \cline{2-3}
    	
				&Quiver Hilbert series &  
$
H_{00}=\frac{1-t^{4(n+1)/3}}{(1-t^{2/3})(1-t^{4/3})(1-t^{2(n+1)/3)})^2}
$\\ \cline{2-3}
				&Single trace index & $\mathcal{I}_{s.t}=\frac{(n+1)t^2}{1-t^2}+2\frac{t^{2n+2}}{1-t^{2n+2}}-\frac{(n+1)(t^2-t^4)}{(1-t^3y^{-1})(1-t^3y)}$ \\
				\hline
				\multirow{3}{*}{\makecell{$D_n$}} &
				Central charges $(a,c)$ & $a=-\frac{5}{24}(n+1)+(n-2)N^2\quad c=-\frac{1}{6}(n+1)+(n-2)N^2$
				 \\ 
				\cline{2-3}
				&Quiver Hilbert series & $H_{00}=\frac{1-t^{8(n-1)/3}}{(1-t^{2/3})(1-t^{8/3})(1-t^{4(n-2)/3})(1-t^{4(n-1)/3})}$  \\  \cline{2-3}
				& Single trace index & $
\mathcal{I}_{s.t.}=\frac{(n+1)t^2}{1-t^2}+\frac{2t^8}{1-t^8}+\frac{t^{4n-8}}{1-t^{4n-8}}-\frac{t^4}{1-t^4}-\frac{(n+1)(t^2-t^4)}{(1-t^3y^{-1})(1-t^3y)}
$\\
				\hline
				\multirow{3}{*}{$E_6$}&
				Central charges $(a,c)$ & $a=-\frac{35}{24}+6N^2\quad c=-\frac{7}{6}+6N^2$
				 \\ 
				\cline{2-3}
				&Quiver Hilbert series &  $\frac{1-t^{16}}{(1-t^{2/3})(1-t^4)(1-t^{16/3})(1-t^8)}$\\ \cline{2-3}
				&Single trace index &$
\mathcal{I}_{s.t.}=\frac{7t^2}{1-t^2}+\frac{t^8}{1-t^8}+\frac{2t^{12}}{1-t^{12}}-\frac{t^4}{1-t^4}-\frac{7(t^2-t^4)}{(1-t^3y^{-1})(1-t^3y)}
$ \\
				\hline
				\multirow{3}{*}{$E_7$}&
				Central charges $(a,c)$ & $a=-\frac{5}{3}+12N^2\quad c=-\frac{4}{3}+12N^2$
				\\ 
				\cline{2-3}
				&Quiver Hilbert series &  $H_{00}=\frac{1-t^{24}}{(1-t^{2/3})(1-t^{16/3})(1-t^8)(1-t^{12})}  $\\ \cline{2-3}
				&Single trace index &$
\mathcal{I}_{s.t.}=\frac{8t^2}{1-t^2}+\frac{t^8}{1-t^8}+\frac{t^{12}}{1-t^{12}}+\frac{t^{16}}{1-t^{16}}-\frac{t^4}{1-t^4}-\frac{8(t^2-t^4)}{(1-t^3y^{-1})(1-t^3y)}
$ \\
				\hline
				\multirow{3}{*}{$E_8$}&
				Central charges $(a,c)$ & $a=-\frac{15}{8}+30N^2\quad c=-\frac{3}{2}+30N^2$
				 \\ 
				\cline{2-3}
				&Quiver Hilbert series & $H_{00}=\frac{1-t^{40}}{(1-t^{2/3})(1-t^8)(1-t^{40/3})(1-t^{20})}$  \\ \cline{2-3}
				&Single trace index & $
\mathcal{I}_{s.t.}=\frac{9t^2}{1-t^2}+\frac{t^8}{1-t^8}+\frac{t^{12}}{1-t^{12}}+\frac{t^{20}}{1-t^{20}}-\frac{t^4}{1-t^4}-\frac{9(t^2-t^4)}{(1-t^3y^{-1})(1-t^3y)}
$ \\
				\hline
			\end{tabular}
		}
	\end{center}
	\caption{Field theory data for affine ADE quiver gauge theory whose dual singularity is $\mathbb{C}^2/G$ with $G=ADE$  finite subgroups of $SU(2)$.}
	\label{quivergaugeSu2}
\end{table}

\newpage
\subsection{Relevant deformations}
Let's now study relevant deformations of the above $ADE$ quiver gauge theory. An obvious choice would be turning on the mass deformation for all the adjoint chiral fields. The IR $R$ charge of all the chiral fields now becomes 
$\frac{1}{2}$, and the corresponding central charges $a,c$, Hilbert series and single trace index are listed in Table \ref{quivergaugeSu2integratedout}. From the Hilbert series, one may conjecture that the dual geometry is given by the hypersurface singularity 
\begin{equation*}
f_{ADE}(x,y,z)+w^h=0,
\end{equation*}
where $h$ is the Coxeter number, and $f_{ADE}$ is the ADE singularity, see Table \ref{su2subgroupwithoutloops}. We also compute the integral second homology group of 
the corresponding link, see Table \ref{tab:deformedade}: $b_2$ is equal to the rank of the type of singularity, and 
it is torsion free. Such deformations for the field theory have already been studied in \cite{Gubser:1998ia}, we put the dual geometry in a more explicit form, and our computation of the Hilbert series and index would confirm the duality. 
Unlike the ADE singularity, our new singularity has the so-called homogeneous deformation (the deformation has the same weight as the original polynomial), for example, the  $A$
type geometry can be deformed as follows:
\begin{equation*}
x^2+y^2+z^n+w^n+\sum_{i=2}^{n-2} \tau_i z^{i} w^{n-i}=0\,.
\end{equation*}
These $\tau$'s are identified with the exact marginal deformations. These singularities also admit
crepant resolutions, and the number of exceptional curves is equal to $\rank(G)$ ($G=ADE$), so noncommutative 
crepant resolutions also exist \cite{van2004non}. The number of gauge groups in the NCCR is equal to $\rank(G)+1$ (the number of exceptional curves plus one), which is exactly as that in our proposed quiver.

\begin{table}[H]
\begin{center}\resizebox{1\columnwidth}{!}{
\begin{tabular}{|c|c|l|c|c|c|} \hline 
Group $G$& Name &  Deformed Equation  &$R$ charge $(d;w_x,w_y,w_z,w_w)$ & Hilbert series $H_{00}$& Leading order of $a,c$\\ \hline 
Cyclic &   $A_{n}$ & $x^2+y^2+z^{n+1}+w^{n+1}=0$   &$\left\{ (n+1);\frac{ (n+1)}{2},\frac{ (n+1)}{2},1,1\right\} $ & $H_{00}=\frac{1-t^{n+1}}{(1-t^{(n+1)/2})(1-t^{(n+1)/2})(1-t)(1-t)}$& $\frac{27}{32}\frac{n+1}{4}N^2$ \\ \hline  
 $\mathbb{D}$ &  $D_{n}$& $x^2+y^{n-1}+yz^2+w^{2n-2}=0$  &$\left\{2(n-1);n-1,2,(n-2),1\right\}$& $H_{00}=\frac{1-t^{2(n-1)}}{(1-t^{n-1})(1-t^{2})(1-t^{n-2})(1-t)}$ &$\frac{27}{32}(n-2)N^2$ \\   \hline 
 $\mathbb{T}$ & $E_6$  & $x^2+y^3+z^4+w^{12}=0$  &  $\left\{12;6,4,3,1\right\}$ & $H_{00}=\frac{1-t^{12}}{(1-t^{6})(1-t^{4})(1-t^3)(1-t)}$&$\frac{27}{32}6N^2$  \\ \hline 
   $\mathbb{O}$ & $E_7$  & $x^2+y^3+yz^3+w^{18=0}$ &  $\left\{18;9,6,4,1\right\}$ & $H_{00}=\frac{1-t^{18}}{(1-t^{9})(1-t^{6})(1-t^4)(1-t)}$  &$\frac{27}{32}12 N^2$ \\ \hline 
 $\mathbb{I}$ & $E_8$  &$x^2+y^3+z^5+w^{30}=0$  &   $\left\{30;15,10,6,1\right\}$  & $H_{00}=\frac{1-t^{30}}{(1-t^{15})(1-t^{10})(1-t^6)(1-t)}$&$\frac{27}{32}30N^2$\\ \hline 
\end{tabular}}
\end{center}
\caption{Geometric data of the corresponding singularity $\mathbb{C}^2/G ( G=ADE)$  whose dual quiver is the affine $ADE$ quiver with adjoint chirals integrated out, and the $R$ charge is found by requiring the top form to have charge $2$.}
\label{su2subgroupwithoutloops}
\end{table}

\textbf{Remark 1}: The deformed superpotential for above theories is $W+\sum m_i \Tr \phi_i^2$, where $\phi_i$'s are the adjoint chiral fields and $W$ is the original superpotential. One may turn on higher order deformation, i.e. $W+\sum m_i \Tr \phi_i^n $, and the added deformation is irrelevant \footnote{When $n=3$, the deformation is marginal irrelevant.}. However, one might still 
find out the dual geometry $f_{ADE}(x,y,z)+w^{(n-1)h}=0$ by computing quiver Hilbert series \cite{Cachazo:2001jy} from using the $R$ charge compatible with superpotential and naive anomaly free condition on $R$ charge.

\textbf{Remark 2}: The deformed geometry $f_{A_n}(x,y,z)+w^{p}=0,~~\frac{n+1}{2}<p<2(n+1)$ is also stable \cite{collins2019sasaki}, but one could not find a quiver gauge theory description for the field theory as the singularity admits no crepant resolution.
These situations would be discussed elsewhere.

\begin{figure}[H]
  \begin{center}
\begin{tikzcd}[ampersand replacement=\&] 
				*\ar[dr,bend left=10,"C"]\ar[dd,bend left=10,"a"]\&\\
				\&N\ar[ul,bend left=10,"c"]\ar[dl,bend left=10,"D"]
				\\
				N\ar[uu,bend left=10,"A"]\ar[ur,bend left=10,"d"]\&\\
    \textbf{Q}_{A2}
			\end{tikzcd}
\begin{tikzcd}\Longrightarrow
\end{tikzcd}
\begin{tikzcd}[ampersand replacement=\&]
	N\ar[dr,bend left=10,"c^*"]\ar[dd,bend left=10,"A^*"]\&\\
				\&N\ar[ul,bend left=10,"C^*"]\ar[dl,bend left=40,"D"]\ar[loop,out=110,in=70,looseness=9,"{[Cc]}"]\ar[dl,bend left=25,"{[ca]}"description]\\
	N\ar[ur,bend left=5,"{[AC]}"description]\ar[loop,out=290,in=250,looseness=13,"{[Aa]}"]\ar[uu,bend left=10,"a^*"]\ar[ur,bend left=20,"d"]\&\\
			\end{tikzcd}
\begin{tikzcd}
    \Longrightarrow
\end{tikzcd}
\begin{tikzcd}[ampersand replacement=\&] 
					N\ar[dr,bend left=10,"C"]\ar[dd,bend left=10,"a"]\&\\
				\&N\ar[ul,bend left=10,"c"]\ar[dl,bend left=10,"D"]
				\\
				N\ar[uu,bend left=10,"A"]\ar[ur,bend left=10,"d"]\&\\
   \textbf{Q}_{A2}
			\end{tikzcd}
    \caption{Seiberg duality of affine $A_2$ quiver.}
    \label{A2seiberg}
    \end{center}
\end{figure}
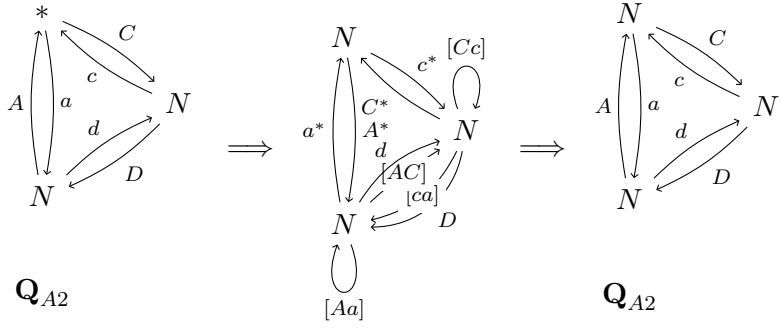

\textbf{Remark 3}: Since the adjoints are integrated out, one can perform the Seiberg duality on these theories, and it enjoys similar self-dual property as studied in conifold example \cite{Klebanov:2000hb}. The following is an example for affine $A_2$ Seiberg duality, see Figure \ref{A2seiberg}.

\begin{table}[H]
    \centering
    \scalebox{0.8}{
    \begin{tabular}{|c|c|c|c|c|}
\hline
Name&Equation &$(u_1,u_2,u_3,u_4;v_1,v_2,v_3,v_4)$& $b_2:$ rank of $H_2(L)$ & $H_2(L)^{tor}$\\
\hline
 $A_n$& $x^2+y^2+z^{n+1}+w^{n+1}=0$&$(n+1,n+1,2,2;1,1,1,1)$&$n$ & No \\
\hline
 $D_n$& $x^2+y^{n-1}+yz^2+w^{2n-2}=0$&\makecell{$(2n-2,n-1,n-1,2;1,1,
 \frac{n-2}{2},1)$ for n even\\ $(2n-2,n-1,2n-2,2;1,1,n-2,1)$ for n odd} &  $n$ & No\\  
\hline
 $E_6$& $x^2+y^3+z^4+w^{12}=0$&$(12,4,3,2;1,1,1,1)$ & 6 & No\\
\hline
 $E_7$& $x^2+y^3+yz^3+w^{18}=0$&$(18,9,3,2;1,2,1,1)$ & 7& No\\
\hline
 $E_8$& $x^2+y^3+z^5+w^{30}=0$&$(30,5,3,2;1,1,1,1)$ &8& No\\
 \hline
    \end{tabular}}
    \caption{The geometrical data for the singularity $f_{ADE}(x,y,z)+w^h=0$, and its dual theory is given by integrating out the adjoints of the affine ADE quiver. Here $\frac{u_i}{v_i}={d\over w_i}$, with $u_i, v_i$ coprime, see Table \ref{su2subgroupwithoutloops} for the weights.}
    \label{tab:deformedade}
\end{table}

\begin{table}[htbp]
	\begin{center}
		\renewcommand\arraystretch{2.2}
		\scalebox{0.8}{
			\begin{tabular}{|l|l|l|} \hline
				Group & Field theory data&\\  \hline
			\multirow{3}{*}{\makecell{ $A_n$}} &
				Central charges $(a,c)$ & 
				$a=\frac{3}{128}(n+1)(9N^2-8)\quad c=\frac{1}{128}(n+1)(27N^2-16)$\\ \cline{2-3}
    	
				&Quiver Hilbert series &  
$
H_{00}=\frac{1-t^{n+1}}{(1-t)^2(1-t^{(n+1)/2})^2}
$\\ \cline{2-3}
				&Single trace index &  $\mathcal{I}_{s.t.}=\frac{(n+1) t^3}{1-t^3}+\frac{2t^{3(n+1)/2}}{1-t^{3(n+1)/2}}$\\
				\hline
				\multirow{3}{*}{\makecell{$D_n$}} &
				Central charges $(a,c)$ & $a=\frac{3}{32}(-2-18N^2+n(9N^2-2))\quad c=\frac{1}{32}(-4-54N^2+n(-4+27N^2))$
				 \\ 
				\cline{2-3}
				&Quiver Hilbert series & $H_{00}=\frac{1-t^{2n-2}}{(1-t)(1-t^2)(1-t^{n-2})(1-t^{n-1})}$  \\  \cline{2-3}
				& Single trace index & $\mathcal{I}_{s.t.}=\frac{nt^3}{1-t^3}+\frac{2t^6}{1-t^6}+\frac{t^{3n-6}}{1-t^{3n-6}}$\\
				\hline
				\multirow{3}{*}{$E_6$}&
				Central charges $(a,c)$ & $a=\frac{3}{16}(-7+27N^2)\quad c=-\frac{7}{8}+\frac{81}{16}N^2$
				 \\ 
				\cline{2-3}
				&Quiver Hilbert series &  $\frac{1-t^{12}}{(1-t)(1-t^3)(1-t^4)(1-t^6)}$\\ \cline{2-3}
				&Single trace index & $\mathcal{I}_{s.t.}=\frac{6t^3}{1-t^3}+\frac{t^6}{1-t^6}+\frac{2t^9}{1-t^9}$\\
				\hline
				\multirow{3}{*}{$E_7$}&
				Central charges $(a,c)$ & $a=\frac{3}{8}(-4+27N^2)\quad c=-1+\frac{81}{8}N^2$ 
				\\ 
				\cline{2-3}
				&Quiver Hilbert series &  $H_{00}=\frac{1-t^{18}}{(1-t)(1-t^4)(1-t^6)(1-t^9)}                                             $ \\ \cline{2-3}
				&Single trace index & $\mathcal{I}_{s.t.}=\frac{7t^3}{1-t^3}+\frac{t^6}{1-t^6}+\frac{t^{9}}{1-t^{9}}+\frac{t^{12}}{1-t^{12}}$ \\
				\hline
				\multirow{3}{*}{$E_8$}&
				Central charges $(a,c)$ & $a=\frac{27}{16}(-1+15N^2)\quad c=\frac{9}{16}(-2+45N^2)$ 
				 \\ 
				\cline{2-3}
				&Quiver Hilbert series & $H_{00}=\frac{1-t^{30}}{(1-t)(1-t^6)(1-t^{10})(1-t^{15})}$  \\ \cline{2-3}
				&Single trace index & $\mathcal{I}_{s.t.}=\frac{8t^3}{1-t^3}+\frac{t^6}{1-t^6}+\frac{t^9}{1-t^9}+\frac{t^{15}}{1-t^{15}}$ \\
				\hline
			\end{tabular}
		}
	\end{center}
	\caption{Field theory data for deformed affine $ADE$ quiver where the adjoint chiral superfields are integrated out.}
	\label{quivergaugeSu2integratedout}
\end{table}

\clearpage
\section{Finite subgroups of $SO(3)$}\label{fintesubgpso3}
In this section, we will discuss the quiver gauge theory corresponding to finite subgroups of $SO(3)$ group and the geometry data is listed in Table \ref{SO3}.

\begin{table}[htbp]
\begin{center}
\resizebox{1\columnwidth}{!}{
\begin{tabular}{|c|c|c|c|c|c|} \hline 
Group $G$& Order $|G|$ & Equation &  \makecell{R charge \\ $(f; x,y,z,w)$} &Hilbert series $H_{00}$& Leading order of $a,c$   \\ \hline
Cyclic $\mathbb{Z}_{n+1}$&n+1& $x^2+y^2+z^{n+1}=0$ & $\left\{\frac{4 (n+1)}{3};\frac{2 (n+1)}{3},\frac{2 (n+1)}{3},\frac{4}{3},\frac{2}{3}\right\} $ & $H_{00}=\frac{1-t^{4(n+1)/3}}{(1-t^{2(n+1)/3})(1-t^{2(n+1)/3})(1-t^{4/3})(1-t^{2/3})}$&$\frac{n+1}{4}N^2$  \\ \hline  
 Dihedral $D_{4m}$& 4m& $x^2+y^{2m+1}+yz^2+wy^{2m}+\ldots=0$&  $\left\{\frac{4}{3} (2 m+1);\frac{2}{3} (2 m+1),\frac{4}{3},\frac{4 m}{3},\frac{4}{3}\right\}$& $H_{00}=\frac{1-t^{4(2m+1)/3}}{(1-t^{2(2m+1)/3})(1-t^{4m/3})(1-t^{4/3})(1-t^{4/3})}$& $mN^2$\\ \hline 
 Dihedral $D_{4m+2}$& 4m+2&$x^2+y^{2m+2}+yz^2+wy^{2m+1}+\ldots=0$ &  $\left\{\frac{8 (m+1)}{3};\frac{4 (m+1)}{3},\frac{4}{3},\frac{2}{3} (2 m+1),\frac{4}{3}\right\}$ &$H_{00}=\frac{1-t^{8(m+1)/3}}{(1-t^{4(m+1)/3})(1-t^{4/3})(1-t^{(4m+2)/3})(1-t^{4/3})}$ &$\frac{1}{2} (1 + 2 m)N^2$\\ \hline
Tetrahedral ${\cal T}$ & 12&$x^2+y^3+z^4+w^2y^2+w^3z^2+w y z^2=0$ &  $(8;4,\frac{8}{3},2,\frac{4}{3})$  & $H_{00}=\frac{1-t^{8}}{(1-t^{4})(1-t^{8/3})(1-t^{2})(1-t^{4/3})}$ &$3N^2$  \\ \hline 
 Octahedral ${\cal O}$ & 24&$x^2+y^3+yz^3+w^3 y^2+\ldots=0$ &  $\left\{12;6,4,\frac{8}{3},\frac{4}{3}\right\}$ & $H_{00}=\frac{1-t^{12}}{(1-t^{6})(1-t^{4})(1-t^{8/3})(1-t^{4/3})}$&$6N^2$\\ \hline 
 Icosahedral ${\cal I}$ & 60&$x^2+y^3+z^5+w^3 z^4+w^5y^2+w^4z^2y+w^2zy^2+wz^3y=0$ &  $\left\{20;10,\frac{20}{3},4,\frac{4}{3}\right\}$ &$H_{00}=\frac{1-t^{20}}{(1-t^{10})(1-t^{20/3})(1-t^4)(1-t^{4/3})}$ &$15N^2$ \\ \hline 
\end{tabular}}
\end{center}
\caption{Geometric data of the corresponding singularity $\mathbb{C}^3/G$ with $G$ finite subgroups of $SO(3)$, and the $R$ charge is determined by requiring the top form to have $R$ charge 2. }
\label{SO3}
\end{table}

For each of these finite subgroups, we have a quiver gauge theory for the corresponding singularity. 
The corresponding McKay quiver and the superpotential are listed in Table \ref{mckaySO3}. We also list the physical quantities including central charges $a,c$, quiver Hilbert series and single trace index for each quiver gauge theory in  Table \ref{quivergaugeSO3}. The expansions of the single trace index are shown in Table \ref{STIexpansion}, which would be useful for the comparison with the computation on gravity side.

For the dihedral group, the quiver is slightly different for the limiting cases. When $n$ is even, the limiting case is $D_4$. When $n$ is odd, the limiting case is $D_2$ and $D_6$. The physical data is shown in Table \ref{limitingdihedral}. From the table, we see that compared with the general case, the constant term of central charges and the coefficient of the last term in the single trace index for $D_2$ are different. And all the physical quantities for $D_4$ and $D_6$ are the same as the general case.

{\small\tabcolsep=3pt
\begin{longtable}{|c|c| }
	\hline
	Group&Mckay quiver and superpotential\\
	\hline
	\endfirsthead
	
	\multicolumn{2}{l}{Continued from previous page}\\
	\hline  Group&McKay quiver and superpotential\\ \hline
	\endhead

	\hline \multicolumn{2}{l}{Continued on next page}\\
	\endfoot
	\endlastfoot
			\multirow{4}{*}{\makecell{Cyclic group \\of order $n+1$}}& \begin{tikzcd}[row sep=tiny, column sep=tiny]
				&N\ar[dl,bend left=10,"b_0"]\ar[dr,bend left=10,"a_n"]\ar[loop,out=110,in=70,looseness=9,"c_0"]&\\[-5pt]
				N\ar[ur,bend left=10,"a_0"]\ar[d,bend left=10,"b_1"]\ar[loop,out=170,in=130,looseness=9,"c_1"]&&N\ar[ul,bend left=10,"b_n"]\ar[d,bend left=10,"a_{n-1}"]\ar[loop,out=50,in=10,looseness=9,"c_n"]\\[+10pt]
				N\ar[dr,bend left=10,"b_2"]\ar[u,bend left=10,"a_1"]\ar[loop,out=230,in=190,looseness=9,"c_2"]&&N\ar[dl,bend left=10,dashed, no head]\ar[u,bend left=10,"b_{n-1}"]\ar[loop,out=350,in=310,looseness=9,"c_{n-1}"]\\[-5pt]
				&N\ar[ul,bend left=10,"a_2"]\ar[loop,out=290,in=250,looseness=12,"c_3"]&\\
			\end{tikzcd}\\
			&$W=\sum_{i=0}^{n-1}(a_ib_ic_{i+1}+c_{i+1}a_{i+1}b_{i+1})+c_0a_0b_0+c_0a_nb_n$\\
			\hline
			\multirow{3}{*}{\makecell{Dihedral group \\of order 2n \\$D_{2n}$ (n=2m even)}} &
			\begin{tikzcd}[column sep=normal,row sep=0.9em] 
				N\ar[dr,bend left=5,"c"]\ar[dd,bend left=5,"a"]&&&&&N\ar[dl,bend left=5,"c'"]\ar[dd,bend left=5,"a'"]\\
				&2N\ar[ul,bend left=5,"C"]\ar[dl,bend left=5,"D_0"]\ar[r,bend left=5,"d_1"]\ar[loop,out=110,in=70,looseness=9,"u_1"]
				&2N\ar[l,bend left=5,"D_1"]\ar[loop,out=110,in=70,looseness=9,"u_2"]\ar[r,dashed,no head]
				&2N\ar[r,bend left=5,"d_{m-2}"]\ar[loop,out=110,in=70,looseness=9,"u_{m-2}"]
				&2N\ar[l,bend left=5,"D_{m-2}"]\ar[ur,bend left=5,"C'"]\ar[dr,bend left=5,"B'"]\ar[loop,out=110,in=70,looseness=9,"u_{m-1}"]&\\
				N\ar[uu,bend left=5,"A"]\ar[ur,bend left=5,"d_0"]&&&&&N\ar[ul,bend left=5,"b'"]\ar[uu,bend left=5,"A'"]\\
			\end{tikzcd}    \\
			&$\begin{aligned}
				W/2=&-ad_0C-cD_0A+u_1D_0d_0+u_1Cc-\sum_{i=1}^{m-2}u_id_iD_i\\
				&+\sum_{i=2}^{m=1}u_iD_{i-1}d_{i-1}-u_{m-1}B'b'-u_{m-1}C'c'-a'b'C'-c'B'A'
			\end{aligned}$ \\
			\hline
			\multirow{2}{*}{\makecell{Dihedral group \\of order 2n \\$D_{2n}$ (n=2m+1 odd)}}&
			\begin{tikzcd}[column sep=normal,row sep=0.8em] 
				N\ar[dr,bend left=5,"C"]\ar[dd,bend left=5,"a"]&&&&\\
				&2N\ar[ul,bend left=5,"c"]\ar[dl,bend left=5,"D_0"]\ar[r,bend left=5,"d_1"]\ar[loop,out=110,in=70,looseness=9,"u_1"]
				&2N\ar[l,bend left=5,"D_1"]\ar[loop,out=110,in=70,looseness=9,"u_2"]\ar[r,dashed,no head]
				&2N\ar[r,bend left=5,"d_{m-1}"]\ar[loop,out=110,in=70,looseness=9,"u_{m-1}"]
				&2N\ar[l,bend left=5,"D_{m-1}"]\ar[loop,out=290,in=250,looseness=13,"v"]\ar[loop,out=110,in=70,looseness=9,"u_m"]\\
				N\ar[uu,bend left=5,"A"]\ar[ur,bend left=5,"d_0"]&&&&\\
			\end{tikzcd}\\
			&$\begin{aligned}W/2=-ad_0C-cD_0A+u_1D_0d_0+u_1Cc-\sum_{i=1}^{m-1} u_id_iD_i+\sum_{i=2}^{m}u_iD_{i-1}d_{i-1}-u_mv^2  \end{aligned}$ \\
			\hline
			\multirow{2}{*}{Tetrahedral group}&
			\begin{tikzcd}[column sep=small,row sep=small] 
				N \ar[dr,bend left=15,"b"]&&N\ar[dl, bend left=15, "c"]\\
				&3N\ar[ul, bend left=15, "B"]\ar[ur, bend left=15, "C"]\ar[d, bend left=15, "A"]\ar[loop,out=355,in=320,looseness=8,"v"]\ar[loop,in=185,out=220,looseness=8,"u"]&\\[+15pt]
				&N\ar[u, bend left=15, "a"]&
			\end{tikzcd}\\
			&$W=uAa+uBb+uCc+u^3+v^3+vAa+vBb+vCc$\\
			\hline
			\multirow{2}{*}{Octahedral group}&
			\begin{tikzcd}[column sep=small,row sep=0.8em] 
				&[+15pt]&&2N\ar[dl,bend left=10,"B"]\ar[dr,bend left=10,"c"]&&[+15pt]&\\[+30pt]
				N\ar[rr,bend left=10,"a"]&&3N\ar[ll,bend left=10,"A"]\ar[rr,bend left=10,"d"]\ar[ur,bend left=10,"b"]\ar[loop,out=290,in=250,looseness=12,"u"]&&3N\ar[ll,bend left=10,"D"]\ar[rr,bend left=10,"e"]\ar[ul,bend left=10,"C"]\ar[loop,out=290,in=250,looseness=12,"v"]&&N\ar[ll,bend left=10,"E"]\\
			\end{tikzcd}\\
			&$\begin{aligned}W/6=&uAa-ubB-udD-1/3u^3+veE-vCc-vDd+1/3v^3\\&+(w^2-w)dCB+(w^2-w)Dbc\end{aligned}$ \\
			\hline
			\multirow{2}{*}{Icosahedral group}&
			\begin{tikzcd}[row sep=tiny, column sep=tiny]
				&[+20pt]&&[+20pt]&&3N\ar[dl,bend left=10,"C"]\ar[dr,bend left=10,"d"]&\\[+30pt]
				N\ar[rr,bend left=10,"A"]&&3N\ar[rr,bend left=10,"B"]\ar[ll,bend left=10,"a"]\ar[loop,out=290,in=250,looseness=12,"u"]&&5N\ar[ll,bend left=10,"b"]\ar[rr,bend left=10,"E"]\ar[ur,bend left=10,"c"]\ar[loop,out=290,in=250,looseness=12,"v"]&&4N\ar[ll,bend left=10,"e"]\ar[ul,bend left=10,"D"]\ar[loop,out=290,in=250,looseness=12,"w"]\\
			\end{tikzcd}\\
			&$\begin{aligned}W/12=&-uAa+5ubB-2/3u^3+15vBb-5vcC-20vEe+10/3v^3\\&+5weE-wDd-1/3w^3-5dec+10CED\end{aligned}$\\
			\hline
			\caption{McKay quivers with superpotentials corresponding to finite subgroups of $SO(3)$.}
			\label{mckaySO3}
\end{longtable}
}

\begin{table}[H]
	\begin{center}
	\renewcommand\arraystretch{2.2}
	\scalebox{0.8}{
		\begin{tabular}{|c|c|c|c|} \hline
		\multirow{3}{*}{$D_4$}& \multirow{3}{*}{\begin{tikzcd}[ampersand replacement=\&] 
					N \ar[r,bend left=10]\ar[d, bend left=10] \ar[dr,bend left=10]\& N  \ar[l,bend left=10]\ar[d, bend left=10]\ar[ld,bend left=10] \\
					N \ar[r,bend left=10]\ar[u, bend left=10]\ar[ur,bend left=10]  \& N \ar[l,bend left=10]\ar[u, bend left=10] \ar[ul,bend left=10]\\
			\end{tikzcd}}
			&Central charges $(a,c)$ & 
			$\begin{aligned}a=-3/4+N^2,\quad c=-1/2+N^2\end{aligned}$ \\ 
			\cline{3-4}
			&&Quiver Hilbert series &  $\begin{aligned} H_{00}=\frac{1-t^{4}}{(1-t^{4/3})^3(1-t^{2})}\end{aligned}$ \\ \cline{3-4}
			&&Single trace index & $\begin{aligned}\mathcal I_{s.t.}=\frac{6t^4}{1-t^4}\end{aligned}$ \\
			\hline
		\multirow{3}{*}{$D_2$}& \multirow{3}{*}{\begin{tikzcd}[ampersand replacement=\&] 
				N \ar[r,bend left=20]\ar[r, bend left=10] \ar[loop,out=170,in=190,looseness=13] \&
				N \ar[l,bend left=20]\ar[l, bend left=10] \ar[loop,out=350,in=10,looseness=13] \\
			\end{tikzcd}}
			&Central charges $(a,c)$ & 
			$\begin{aligned}a=-5/12+1/2N^2,\quad c=-1/3+1/2N^2\end{aligned}$ \\ 
			\cline{3-4}
			&&Quiver Hilbert series &  $\begin{aligned} H_{00}=\frac{1-t^{8/3}}{(1-t^{4/3})^3(1-t^{2/3})}\end{aligned}$ \\ \cline{3-4}
			&&Single trace index & $\begin{aligned}\mathcal  I_{s.t.}=\frac{2t^2}{1-t^2}+\frac{2t^4}{1-t^4}-\frac{2(t^2-t^4)}{(1-yt^3)(1-y^{-1}t^3)}\end{aligned}$ \\
			\hline
		\multirow{3}{*}{$D_6$}& \multirow{3}{*}{\begin{tikzcd}[ampersand replacement=\&] 
				N\ar[dr,bend left=10]\ar[dd,bend left=10]\&\\
				\&2N\ar[ul,bend left=10]\ar[dl,bend left=10]\ar[loop,out=110,in=70,looseness=9]\ar[loop,out=290,in=250,looseness=13]
				\\
				N\ar[uu,bend left=10]\ar[ur,bend left=10]\&\\
			\end{tikzcd}}
			&Central charges $(a,c)$ & 
			$\begin{aligned}a=-29/48+3/2N^2,\quad c=-11/24+3/2N^2\end{aligned}$ \\ 
			\cline{3-4}
			&&Quiver Hilbert series &  $\begin{aligned}H_{00}=\frac{1-t^{16/3}}{(1-t^{4/3})^2(1-t^{2})(1-t^{8/3})}\end{aligned}$ \\ \cline{3-4}
			&&Single trace index & $\begin{aligned}\mathcal  I_{s.t.}=\frac{2t^2}{1-t^2}+\frac{2t^4}{1-t^4}+\frac{t^6}{1-t^6}-\frac{2(t^2-t^4)}{(1-yt^3)(1-y^{-1}t^3)}\end{aligned}$ \\
			\hline
		\end{tabular}}
	\end{center}
	\caption{Field theory data for quiver theories whose dual singularity is $\mathbb{C}^3/G$ with G the Limiting cases of the dihedral groups.}
	\label{limitingdihedral}
\end{table}

\begin{table}[H]
	\begin{center}
		\renewcommand\arraystretch{2.2}
		\scalebox{0.8}{
			\begin{tabular}{|l|l|l|} \hline
				Group & Field theory data&\\  \hline
				\multirow{3}{*}{\makecell{Cyclic group \\of order $n+1$}} &
				Central charges $(a,c)$ & 
				$a=\frac{1}{4}(n+1)N^2-\frac{5}{24}(n+1),\quad c=\frac{1}{4}(n+1)N^2-\frac{1}{6}(n+1)$\\ \cline{2-3}
				&Quiver Hilbert series &  $H_{00}=\frac{1-t^{(4n+4)/3}}{(1-t^{2/3})(1-t^{4/3})(1-t^{(2n+2)/3})^2}$ \\ \cline{2-3}
				&Single trace index & $ \mathcal I_{s.t.}=(n+1)\frac{t^2}{1-t^2}+2\frac{t^{2+2n
				}}{1-t^{2+2n}}-(n+1)\frac{t^2-t^4}{(1-y^{-1}t^3)(1-yt^3)}$  \\
				\hline
				\multirow{3}{*}{\makecell{Dihedral group \\of order 2n \\$D_{2n}$ (n=2m even)}} &
				Central charges $(a,c)$ & 
				$\begin{aligned}a=-\frac{13}{24} + m (-\frac{5}{24}+ N^2),\quad c=-\frac{1}{3} + m (-\frac{1}{6}+ N^2)\end{aligned}$ \\ 
				\cline{2-3}
				&Quiver Hilbert series &  $\begin{aligned} H_{00}=\frac{1-t^{(8m+4)/3}}{(1-t^{4/3})^2(1-t^{4m/3})(1-t^{1-t^{(4m+2)/3}})}\end{aligned}$ \\ \cline{2-3}
				&Single trace index & $\begin{aligned} \mathcal I_{s.t.}=(m-1)\frac{t^2}{1-t^2}+5\frac{t^4}{1-t^4}+\frac{t^{4m}}{1-t^{4m}}-(m-1)\frac{t^2-t^4}{(1-y^{-1}t^3)(1-yt^3)}\end{aligned}$ \\
				\hline
				\multirow{3}{*}{\makecell{Dihedral group \\of order 2n \\$D_{2n}$ (n=2m+1 odd)}}&
				Central charges $(a,c)$ & 
				$\begin{aligned}a=-\frac{19}{48}  -\frac{5}{24}m+ N^2(m+\frac{1}{2}),\quad c=-\frac{7}{24} -\frac{1}{6}m+N^2(m+\frac{1}{2})\end{aligned}$ \\ 
				\cline{2-3}
				&Quiver Hilbert series &  $\begin{aligned} H_{00}=\frac{1-t^{(8m+8)/3}}{(1-t^{4/3})^2(1-t^{(4m+2)/3})(1-t^{(4m+4)/3})}\end{aligned}$ \\ \cline{2-3}
				&Single trace index & $\begin{aligned} \mathcal I_{s.t.}=(m+1)\frac{t^2}{1-t^2}+2\frac{t^4}{1-t^4}+\frac{t^{4m+2}}{1-t^{4m+2}}-(m+1)\frac{t^2-t^4}{(1-y^{-1}t^3)(1-yt^3)} \end{aligned}$ \\
				\hline
				\multirow{3}{*}{Tetrahedral group}&
				Central charges $(a,c)$ & 
				$\begin{aligned}a=-\frac{19}{24} + 3N^2,\quad c=-\frac{7}{12} + 3N^2\end{aligned}$ \\ 
				\cline{2-3}
				&Quiver Hilbert series &  $\begin{aligned} H_{00}=\frac{1 - t^8}{(1 - t^\frac{4}{3}) (1 - t^2) (1 - t^\frac{8}{3}) (1 - t^4)}\end{aligned}$ \\ \cline{2-3}
				&Single trace index & $\begin{aligned} \mathcal I_{s.t.}=2\frac{t^2}{1-t^2}+2\frac{t^4}{1-t^4}+2\frac{t^6}{1-t^6}-2\frac{t^2-t^4}{(1-t^3y^{-1})(1-t^3y)}\end{aligned}$ \\
				\hline
				\multirow{3}{*}{Octahedral group}&
				Central charges $(a,c)$ & 
				$\begin{aligned}a=-\frac{47}{48} + 6N^2,\quad c=-\frac{17}{24} + 6N^2 \end{aligned}$ \\ 
				\cline{2-3}
				&Quiver Hilbert series &  $\begin{aligned} H_{00}=\frac{1 - t^{12}}{(1 - t^\frac{4}{3})  (1 - t^\frac{8}{3}) (1 - t^4)(1 - t^6)}\end{aligned}$ \\ \cline{2-3}
				&Single trace index & $\begin{aligned}  \mathcal I_{s.t.}=2\frac{t^2}{1-t^2}+3\frac{t^4}{1-t^4}+\frac{t^{6}}{1-t^{6}}+\frac{t^{8}}{1-t^{8}}-2\frac{t^2-t^4}{(1-y^{-1}t^3)(1-yt^3)}\end{aligned}$ \\
				\hline
				\multirow{3}{*}{Icosahedral group}&
				Central charges $(a,c)$ & 
				$\begin{aligned}a=-1+ 15N^2,\quad c=-\frac{3}{4} + 15N^2 \end{aligned}$ \\ 
				\cline{2-3}
				&Quiver Hilbert series &  $\begin{aligned} H_{00}=\frac{1-t^{20}}{(1-t^{4/3})(1-t^4)(1-t^{20/3})(1-t^{10})}\end{aligned}$ \\ \cline{2-3}
				&Single trace index & $\begin{aligned}  \mathcal I_{s.t.}=3\frac{t^2}{1-t^2}+2\frac{t^4}{1-t^4}+\frac{t^{6}}{1-t^{6}}+\frac{t^{10}}{1-t^{10}}-3\frac{t^2-t^4}{(1-y^{-1}t^3)(1-yt^3)}\end{aligned}$ \\
				\hline
			\end{tabular}
		}
	\end{center}
	\caption{Field theory data for quiver gauge theories whose dual singularity is $\mathbb{C}^3/G$ with $G$ finite subgroups of $SO(3)$.}
	\label{quivergaugeSO3}
\end{table}

\clearpage

\begin{table}[htp]
	\begin{center}
		\scalebox{0.85}{
		\begin{tabular}{|c|c|} \hline
  Group& Single trace expansion\\
  \hline
  \makecell{Cylic group \\of order $n+1$}& $
\begin{aligned}
\mathcal{I}_{s.t}&=\frac{1}{(1-y^{-1}t^3)(1-y^{1}t^3)}(-(n+1)t^2+(n+1)t^4+(n+1)t^2(1+t^6)\sum_{n=0}^{\infty}t^{2n}+\\&2t^{2+2n}(1+t^6)\sum_{s=0}^{\infty}t^{(2+2n)s}-(n+1)t^5\sum_{n=0}^{\infty}t^{2n}(y+y^{-1})-2t^{5+2n}\sum_{s=0}^{\infty}t^{(2+2n)s}(y+y^{-1}))\\
&=-(n+1)\frac{t^2-t^4}{(1-y^{-1}t^3)(1-yt^3)}+(n+1)t^2\sum_{s=0}^{\infty} t^{2s}+2 t^{2+2n}\sum_{s=0}^{\infty}t^{(2+2n)s}=(2+2n)t^4+\cdots\end{aligned}$\\
  \hline
\makecell{Dihedral group \\of order 2n \\$D_{2n}$ (n=2m even)} &$
\begin{aligned}
\mathcal{I}_{s.t.}=&\frac{1}{(1-y^{-1}t^3)(1-y^{1}t^3)}(-(m-1)t^2+(m-1)t^4+(m-1)t^2(1+t^6)\sum_{n=0}^{\infty}t^{2n}\\&+5t^4(1+t^6)\sum_{n=0}^{\infty}t^{4n}+t^{4m}(1+t^6)\sum_{n=0}^{\infty}t^{4mn}
-(m-1)t^5\sum_{n=0}^{\infty}t^{2n}(y+y^{-1})\\&-5t^7\sum_{n=0}^{\infty}t^{4n}(y+y^{-1})-t^{4m+3}\sum_{n=0}^{\infty}t^{4mn}(y+y^{-1}))
\end{aligned}$ \\
			\hline
			\makecell{Dihedral group \\of order 2n \\$D_{2n}$ (n=2m+1 odd)}&
			$\begin{aligned}
			\mathcal{ I}_{s.t.}
			&=\frac{1}{(1-y^{-1}t^3)(1-y^{1}t^3)}(-(m+1)t^2+(m+1)t^4+(m+1)t^2(1+t^6)\sum_{n=0}^{\infty}t^{2n}\\&+2t^4(1+t^6)\sum_{n=0}^{\infty}t^{4n}+t^{4m+2}(1+t^6)\sum_{n=0}^{\infty}t^{(4m+2)n}-(m+1)t^5\sum_{n=0}^{\infty}t^{2n}(y+y^{-1})\\&-2t^7\sum_{n=0}^{\infty}t^{4n}(y+y^{-1})-t^{4m+5}\sum_{n=0}^{\infty}t^{4m+2
				n}(y+y^{-1}))
			\end{aligned}	$\\
			\hline
			Tetrahedral group&$\begin{aligned}
			\mathcal{I}_{s.t.}&=\frac{1}{(1-y^{-1}t^3)(1-y^{1}t^3)}(-2t^2+2t^4+2t^2(1+t^6)\sum_{n=0}^{\infty}t^{2n}+2t^4(1+t^6)\sum_{n=0}^{\infty}t^{4n}\\&+2t^{6}(1+t^6)\sum_{n=0}^{\infty}t^{6n}-2t^5\sum_{n=0}^{\infty}t^{2n}(y+y^{-1})-2t^7\sum_{n=0}^{\infty}t^{4n}(y+y^{-1})-2t^{9}\sum_{n=0}^{\infty}t^{6
				n}(y+y^{-1}))\end{aligned}$
			\\
			\hline
			Octahedral group&
			$\begin{aligned} \mathcal{I}_{s.t.}&=\frac{1}{(1-y^{-1}t^3)(1-y^{1}t^3)}(-2t^2+2t^4+2t^2(1+t^6)\sum_{n=0}^{\infty}t^{2n}+3t^4(1+t^6)\sum_{n=0}^{\infty}t^{4n}\\&+t^{6}(1+t^6)\sum_{n=0}^{\infty}t^{6n}+t^{8}(1+t^6)\sum_{n=0}^{\infty}t^{8n}-2t^5\sum_{n=0}^{\infty}t^{2n}(y+y^{-1})-3t^7\sum_{n=0}^{\infty}t^{4n}(y+y^{-1})\\&-t^{9}\sum_{n=0}^{\infty}t^{6
				n}(y+y^{-1})-t^{11}\sum_{n=0}^{\infty}t^{8
				n}(y+y^{-1})) \end{aligned}$
			\\
			\hline
			Icosahedral group&	$\begin{aligned} \mathcal{I}_{s.t.}&=\frac{1}{(1-y^{-1}t^3)(1-y^{1}t^3)}(-3t^2+3t^4+3t^2(1+t^6)\sum_{n=0}^{\infty}t^{2n}+2t^4(1+t^6)\sum_{n=0}^{\infty}t^{4n}\\&
        +t^6(1+t^6)\sum_{n=0}^{\infty}t^{6n}+t^{10}(1+t^6)\sum_{n=0}^{\infty}t^{10n}
   -3t^5\sum_{n=0}^{\infty}t^{2n}(y+y^{-1})-2t^7\sum_{n=0}^{\infty}t^{4n}(y+y^{-1})\\
   &-t^{9}\sum_{n=0}^{\infty}t^{6
				n}(y+y^{-1})-t^{13}\sum_{n=0}^{\infty}t^{10
				n}(y+y^{-1})) \end{aligned}$
			\\
			\hline
		\end{tabular}}
	\end{center}
	\caption{Expansion of single trace index for theories defined by finite subgroups of $SO(3)$}
	\label{STIexpansion}
\end{table}

\subsection {Relevant deformations}
Let's now study the relevant deformations of those quiver gauge theories. For quiver gauge theory corresponding to 
Dihedral group of order $2n$ ($n=2m$ even) $D_{4n}$, one can turn on mass deformation for all the adjoints, and this forces 
the $R$ charges to be $R(A)=R(a)=R(A^{\prime})=R(a^{\prime})=1$, so they can all be integrated out. 
Other chiral fields all have $R$ charge $\frac{1}{2}$. The IR theory is just the same as those defined by 
 affine D type quiver gauge theory $D_{n+2}$ with adjoint chirals integrated out.

For the Dihedral group of order $2n$ ($n=2m+1$ odd), one can also add mass terms to the adjoints (except $v$), then the $R$ charges for fields $A,a$ must be $R(a)=R(A)=1$, which means that they become massive.
The new quiver is shown in the Figure \ref{D_{2n}massdeformation}, and the R charge for the chiral fields all equal to $\frac{1}{2}$.
	\begin{figure}[H]
	    \centering
	\begin{tikzcd}[row sep=large, column sep=large]
		N\ar[dr,bend left=10,"c"]&&&&\\
		&2N\ar[ul,bend left=10,"C"]\ar[dl,bend left=10,"D_0"]\ar[r,bend left=10,"d_1"]
		&2N\ar[l,bend left=10,"D_1"]\ar[r,dashed,no head]
		&2N\ar[r,bend left=10,"d_{m-1}"]
		&2N\ar[l,bend left=10,"D_{m-1}"]\ar[loop,out=290,in=250,looseness=13,"v"]\\
		N\ar[ur,bend left=10,"d_0"]&&&&\\
	\end{tikzcd}
	\caption{Quiver for Dihedral group $D_{2n}$ of order $2n$ ($n=2m+1$ odd) with all adjoints except $v$ integrated out.}
	\label{D_{2n}massdeformation}
 \end{figure}
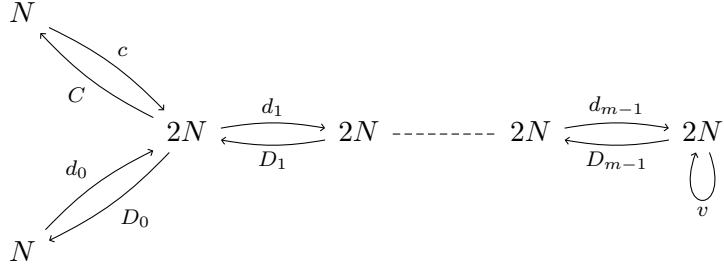
The quiver Hilbert series and the single trace index (here we just subtract the $U(1)$ vector multiplets for the $U(N)$ gauge group) are
\begin{equation}
    \begin{aligned}
            H_{00}&=\frac{1-t^{2m+3}}{(1-t)(1-t^2)(1-t^{(2m+1)/2})(1-t^{(2m+3)/2})},\\
        \mathcal I_{s.t.}&=\frac{(m+1)t^3}{1-t^3}+\frac{t^6}{1-t^6}+\frac{t^{3/2}}{1-t^{3/2}}+\frac{t^{3/2+3m}}{1-t^{3/2+3m}}\,.
    \end{aligned}
    \label{deformindex}
\end{equation}
The central charges are
$$a=\frac{27}{64}(2m+1)N^2-\frac{99+48m}{256},~
c=\frac{27}{64}(2m+1)N^2-\frac{75+32m}{256}\,.$$

From the index form, we see that there is a chiral field with $R$ charge $\frac{1}{2}$ that violates the unitarity bound (this field is easily to be identified with the field $\Tr v$). This field becomes free in the IR, and is decoupled \cite{Kutasov:2003iy}. The correct $U(1)_R$ charge should be the original one mixed with the one acting only on the decoupled field, so that the $R$ charge for the free field is $\frac{2}{3}$. 
The index for the interacting theory might be modified as follows:  subtract the contribution of the free chiral with the prescribed $R$ charge $\frac{1}{2}$, so the index for the interacting theory might be:
\begin{equation}
    \begin{aligned}
        \mathcal I_{s.t.}&=\frac{(m+1)t^3}{1-t^3}+\frac{t^6}{1-t^6}+\frac{t^{3/2}}{1-t^{3/2}}+\frac{t^{3/2+3m}}{1-t^{3/2+3m}}-\frac{t^{3/2}-t^{9/2}}{(1-t^3y)(1-t^3y^{-1})}\,.
    \end{aligned}
\end{equation}
It would be interesting to verify our proposal using other methods.

Based on the quiver Hilbert series, we'd like to conjecture that the dual geometry is:
\begin{equation*}
\boxed{w^2+x^{2m+3}+xy^{m+1}+yz^2=0}\,.
\end{equation*}

\subsection{A detailed study of Seiberg duality}
The exciting feature of the quivers in Figure \ref{mckaySO3} (compare the affine ADE quiver) is that there are
quiver nodes without adjoint chirals, and so one can perform Seiberg duality to get many dual quiver descriptions.

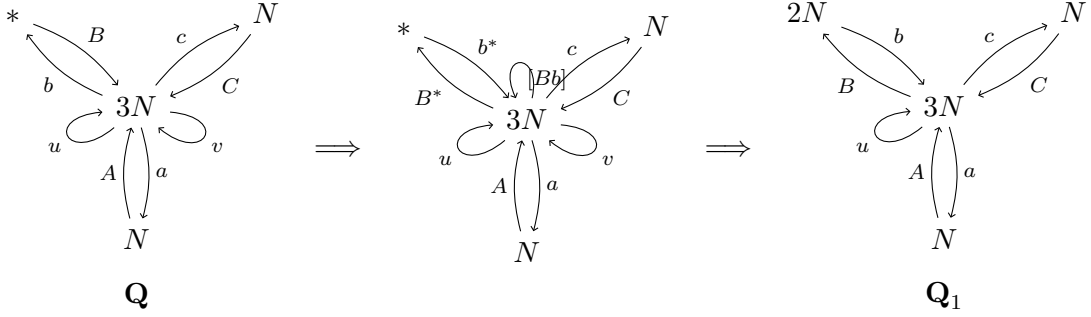
\begin{figure}[H]
	\centering
	\begin{tikzcd}[row sep=normal, column sep=normal]
		* \ar[dr,bend left=15,"B"]&&N\ar[dl, bend left=15, "C"]\\
		&3N\ar[ul, bend left=15, "b"]\ar[ur, bend left=15, "c"]\ar[d, bend left=15, "a"]\ar[loop,in=185,out=220,looseness=8,"u"]\ar[loop,out=355,in=320,looseness=8,"v"]&\\[+15pt]
		&N\ar[u, bend left=15, "A"]&\\[-15pt]
		&\textbf{Q}&
	\end{tikzcd}  
	\begin{tikzcd}
		\stackrel{}{\Longrightarrow}    
	\end{tikzcd}
	\begin{tikzcd}[row sep=normal, column sep=normal]
		* \ar[dr,bend left=15,"b^*"]&&N\ar[dl, bend left=15, "C"]\\
		&3N\ar[ul, bend left=15, "B^*"]\ar[ur, bend left=15, "c"]\ar[d, bend left=15, "a"]\ar[loop,in=185,out=220,looseness=8,"u"]\ar[loop,out=355,in=320,looseness=8,"v"]\ar[loop,in=115,out=78,looseness=8,"{[Bb]}"]&\\[+15pt]
		&N\ar[u, bend left=15, "A"]&\\[-15pt]
		&\textbf{ }&
	\end{tikzcd}  
	\begin{tikzcd}
		\stackrel{}{\Longrightarrow}    
	\end{tikzcd}
	\begin{tikzcd}[row sep=normal, column sep=normal]
		2N \ar[dr,bend left=15,"b"]&&N\ar[dl, bend left=15, "C"]\\
		&3N\ar[ul, bend left=15, "B"]\ar[ur, bend left=15, "c"]\ar[d, bend left=15, "a"]\ar[loop,in=185,out=220,looseness=8,"u"]&\\[+15pt]
		&N\ar[u, bend left=15, "A"]&\\[-15pt]
		&\textbf{Q}_1&
	\end{tikzcd}  
	\caption{Seiberg duality of the McKay quiver $Q$ of the Tetrahedral group.}
        \label{sdtetrahedral}
\end{figure}

Let's discuss the Seiberg duality of theories corresponding to the Tetrahedron group in more detail.
Let's first do Seiberg duality on node $*$ in Figure \ref{sdtetrahedral}. 

The initial superpotential is (the coefficient is chosen so that superpotential for the following duality frame is generic):
\begin{equation}
    \begin{aligned}
        W_{Q}&=uAa+\omega uBb+\omega^2 uCc-\frac{1}{3}u^3-vAa-\omega^2vBb-\omega vCc+\frac{1}{3}v^3,
    \end{aligned}
\end{equation}
where  $\omega=e^{2\pi i/3}$. The new superpotential is
$$
W_{Q_1}=[W_Q]+\Delta=uAa+\omega u[Bb]+\omega^2 uCc-\frac{1}{3}u^3-vAa-\omega^2v[Bb]-\omega vCc+\frac{1}{3}v^3\,
+[Bb]B^*b^*
$$
The reduction is done by integrating out massive superfields $[Bb],v$, namely computing the partial derivative:
\begin{equation}
    \begin{aligned}
\frac{\partial W_{Q_1}}{\partial [Bb]}&=\omega u-\omega^2 v+B^*b^*=0,\\
\frac{\partial W_{Q_1}}{\partial v}&=-Aa-\omega^2[Bb]-\omega Cc+v^2=0\,.
    \end{aligned}
\end{equation}
Thus we get
$$
\boxed{W_{Q_1}=(1-\omega^2)Aau-wAaBb+(\omega^2-1)Ccu-\omega^2 CcbB+\omega^2Bbu^2+\omega (Bb)^2 u+\frac{1}{3}(Bb)^3}
$$
The quiver $Q_1$ is quite different from the original quiver: a) The $U(1)_R$ charges of chiral fields in the IR limit are no longer free: $R(B)=R(b)=\frac{1}{3}$, $R(C)=R(c)=R(A)=R(a)=\frac{2}{3}$, $R(u)=\frac{2}{3}$. 
We compute the Hilbert series $H_{00}$ and the index by using these new $R$ charges, and they give the same 
result as that of the quiver $Q$. The matrix $\chi(t)$ used in computing the index is 
\begin{equation*}
\chi_{Q_{1}}(t)=\left[\begin{array}{cccc}
0&0&t&0 \\
0&0&t^2&0 \\
t&t^2&t^2 &t^2\\
0&0&t^2&0 \\
\end{array}\right]
\end{equation*}

Let's point out a quite interesting point: in the original quiver $Q$, one can see two free chirals as the traceless part of two adjoint chirals
on the middle quiver node; however, the dual theory has just one adjoint chiral whose traceless part would decouple, and the other free chiral is the composite operator $Tr(Bb)$ whose 
$R$ charge is $\frac{2}{3}$ and become decoupled in the IR. This is the typical thing happening in Seiberg duality: the elementary field in one description becomes the composite field 
in another dual description.

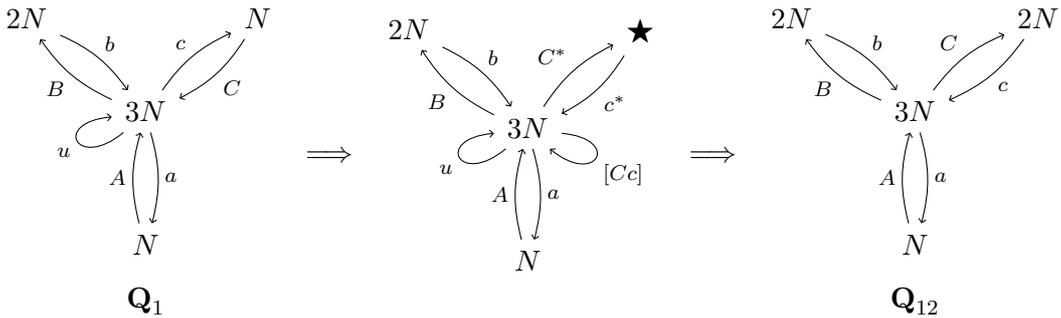
\begin{figure}[H]
	\centering
	\begin{tikzcd}[row sep=normal, column sep=scriptsize]
		2N \ar[dr,bend left=15,"b"]&&N\ar[dl, bend left=15, "C"]\\
		&3N\ar[ul, bend left=15, "B"]\ar[ur, bend left=15, "c"]\ar[d, bend left=15, "a"]\ar[loop,in=185,out=220,looseness=8,"u"]&\\[+15pt]
		&N\ar[u, bend left=15, "A"]&\\[-15pt]
		&\textbf{Q}_1&
	\end{tikzcd}  
	\begin{tikzcd}
		\stackrel{}{\Longrightarrow}
	\end{tikzcd}
	\begin{tikzcd}[row sep=normal, column sep=scriptsize]
		2N \ar[dr,bend left=15,"b"]&&\bigstar\ar[dl, bend left=15, "c^*"]\\
		&3N\ar[ul, bend left=15, "B"]\ar[ur, bend left=15, "C^*"]\ar[d, bend left=15, "a"]\ar[loop,in=185,out=220,looseness=8,"u"]\ar[loop,out=355,in=320,looseness=8,"{[Cc]}"]&\\[+15pt]
		&N\ar[u, bend left=15, "A"]&\\[-15pt]
		&\textbf{}&
	\end{tikzcd}  
	\begin{tikzcd}
		\stackrel{}{\Longrightarrow}
	\end{tikzcd}
	\begin{tikzcd}[row sep=normal, column sep=scriptsize]
		2N \ar[dr,bend left=15,"b"]&&2N\ar[dl, bend left=15, "c"]\\
		&3N\ar[ul, bend left=15, "B"]\ar[ur, bend left=15, "C"]\ar[d, bend left=15, "a"]&\\[+15pt]
		&N\ar[u, bend left=15, "A"]&\\[-15pt]
		&\textbf{Q}_{12}&
	\end{tikzcd}
	\caption{Seiberg duality of the quiver $Q_{1}$.}
	\label{Q1}
\end{figure}

Let's now do Seiberg duality on node $\bigstar$ in the quiver $Q_1$, see Figure \ref{Q1}. The final superpotential after the reduction is 
$$
\boxed{W_{Q_{12}}=AaBb+AaCc-(Bb)^2Cc-Bb(Cc)^2}.
$$
The new $R$ charges would be $R(B)=R(b)=R(C)=R(c)=\frac{1}{3}$, and $R(A)=R(a)=\frac{2}{3}$. Again, one can compute the Hilbert series and index of quiver $Q_{12}$, which are the same as those of $Q$. 
The matrix $\chi(t)$ used in computing the index is 
\begin{equation*}
\chi_{Q_{12}}(t)=\left[\begin{array}{cccc}
0&0&t&0 \\
0&0&t&0 \\
t&t&0 &t^2\\
0&0&t^2&0 \\
\end{array}\right]
\end{equation*}

\begin{figure}[H]
	\centering
	\begin{tikzcd}[row sep=normal, column sep=small]
		2N \ar[dr,bend left=15,"b"]&&2N\ar[dl, bend left=15, "c"]\\
		&3N\ar[ul, bend left=15, "B"]\ar[ur, bend left=15, "C"]\ar[d, bend left=15, "a"]&\\[+15pt]
		&N\ar[u, bend left=15, "A"]&\\
		&\textbf{Q}_{12}&
	\end{tikzcd}
	\begin{tikzcd}
	\stackrel{}{\Longrightarrow}
	\end{tikzcd}
	\begin{tikzcd}[row sep=normal, column sep=normal]
		2N \ar[dr,bend left=10,"B^*"]\ar[loop,out=120,in=70,looseness=6,"u_2"]\ar[rr,bend left=5,"d"]\arrow[to=3-2, bend left=5,"E"]&&2N\ar[dl, bend left=10, "C^*"]\ar[loop,out=120,in=70,looseness=6,"u_3"]\ar[ll, bend left=5,"D"]\arrow[to=3-2, bend left=5,"F"]\\
		& \clubsuit \ar[ul, bend left=15, "b^*"]\ar[ur, bend left=10, "c^*"]\ar[d, bend left=10, "A^*"]&\\[+15pt]
		&N\ar[u, bend left=10, "a^*"]\ar[loop,out=290,in=250,looseness=8,"u_0"]\arrow[to=1-1,bend left=10,"e"]\arrow[to=1-3,bend left=10,"f"]&\\
		&\textbf{}&
	\end{tikzcd} 
	\begin{tikzcd}
		\stackrel{}{\Longrightarrow}
	\end{tikzcd}
	\begin{tikzcd}[row sep=normal, column sep=small]
		2N \ar[dr,bend left=15,"B"]\ar[loop,out=120,in=70,looseness=6,"u_2"]\ar[rr,bend left=10,"d"]
		&&2N\ar[dl, bend left=15, "C"]\ar[loop,out=120,in=70,looseness=6,"u_3"]\ar[ll, bend left=10,"D"]\\
		& 2N \ar[ul, bend left=15, "b"]\ar[ur, bend left=15, "c"]\ar[d, bend left=15, "A"]&\\[+15pt]
		&N\ar[u, bend left=15, "a"]\ar[loop,out=290,in=250,looseness=8,"u_0"]&\\
		&\textbf{Q}_{123}&
	\end{tikzcd}   
	\caption{Seiberg duality of the quiver $Q_{12}$.}
	\label{Q12}
\end{figure}
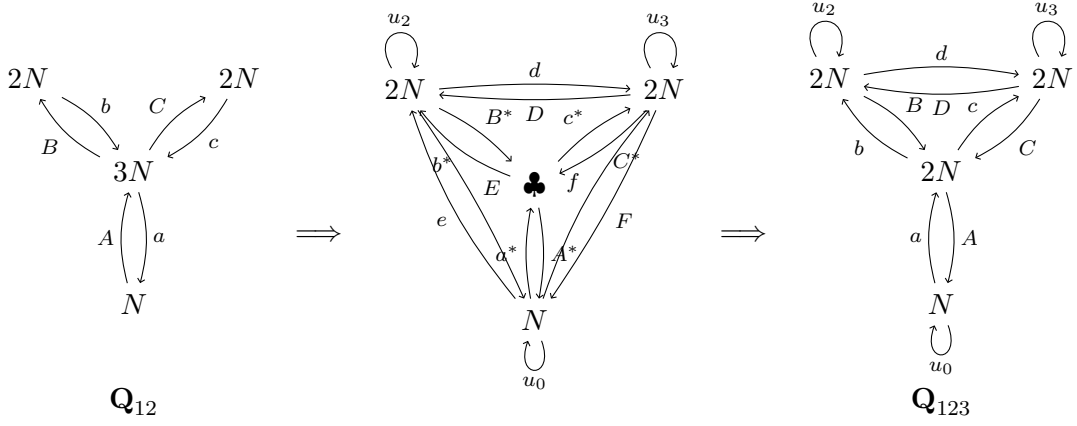

Finally, Let's do Seiberg duality on the central quiver node ($\clubsuit$ node) of $Q_{12}$.  We show the process in Figure \ref{Q12}.
The final superpotential is 
$$
\boxed{W_{Q_{123}}=aAu_0-AaBb-AaCc+bBu_2-dDu_2+cCu_3-Ddu_3+Bdc+CDb}
$$
The $R$ charges are $R(B)=R(b)=R(C)=R(c)=R(D)=R(d)=\frac{2}{3}, R(u_2)=R(u_3)=\frac{2}{3}$, 
$R(A)=R(a)=\frac{1}{3},~R(u_0)=\frac{4}{3}$, which would give the same Hilbert series and index as those of the original quiver.  The matrix $\chi(t)$ used in computing the index is 
\begin{equation*}
\chi_{Q_{123}}(t)=\left[\begin{array}{cccc}
t^2&t^2&t^2&0 \\
t^2&t^2&t^2&0 \\
t^2&t^2&0 &t\\
0&0&t&t^4 \\
\end{array}\right]
\end{equation*}

\begin{figure}[H]
\centering
\begin{tikzcd}[ampersand replacement=\&] 
				N\ar[dr,bend left=10,"C"]\ar[dd,bend left=10,"a"]\&\\
				\&2N\ar[ul,bend left=10,"c"]\ar[dl,bend left=10,"D"]\ar[loop,out=110,in=70,looseness=9,"u"]\ar[loop,out=290,in=250,looseness=13,"v"]
				\\
				N\ar[uu,bend left=10,"A"]\ar[ur,bend left=10,"d"]\&\\
    \textbf{Q}_{D_6}
			\end{tikzcd}
\begin{tikzcd}\Longrightarrow
\end{tikzcd}
\begin{tikzcd}[ampersand replacement=\&] 
	2N\ar[dr,bend left=10,"c^*"]\ar[dd,bend left=10,"A^*"]\&\\
				\&2N\ar[ul,bend left=10,"C^*"]\ar[dl,bend left=10,"D"]\ar[loop,out=110,in=70,looseness=9,"u"]\ar[loop,out=290,in=250,looseness=13,"v"]\ar[loop,out=50,in=10,looseness=9,"{[Cc]}"]\ar[dl,bend left=20,"{[ca]}"]\\
	N\ar[ur,bend left=20,"{[AC]}"]\ar[loop,out=290,in=250,looseness=13,"{[Aa]}"]\ar[uu,bend left=10,"a^*"]\ar[ur,bend left=10,"d"]\&\\
			\end{tikzcd}
\begin{tikzcd}
    \Longrightarrow
\end{tikzcd}
\begin{tikzcd}[ampersand replacement=\&] 
				2N\ar[dr,bend left=10,"C"]\ar[dd,bend left=10,"a"]\&\\
				\&2N\ar[ul,bend left=10,"c"]\ar[loop,out=110,in=70,looseness=9,"{v}"]
				\\
				N\ar[uu,bend left=10,"A"]\ar[loop,out=290,in=250,looseness=9,"{[Aa]}"]\&\\
    \textbf{Q}_{D_6}^1
			\end{tikzcd}
    \caption{Seiberg duality of $D_6$ with the upper node $N$}
    \label{SeibergdualD6}
\end{figure}
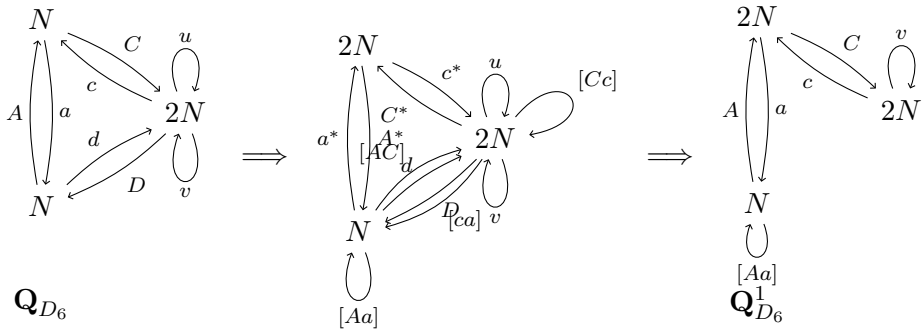

Next we discuss the Seiberg duality of $D_6$ quiver. We can do the Seiberg duality on node with rank $N$, see figure. \ref{SeibergdualD6}.

The superpotential of $D_6$ is
\begin{equation}
    W_{D_6}=ACD+acd+uCc+uDd+uv^2\,.
\end{equation}
The new superpotential is 
\begin{equation}
    W_{D_6}^{\prime}=[W_{D_6}]+\Delta=[AC]D+[ac]d+u[Cc]+uDd+uv^2+[AC]C^*A^*+[ac]c^*a^*+[Cc]c^*C^*+[Aa]a^*A^*\,.
\end{equation}
We can do reduction by integrating out massive superfields $[AC],D,[ac],d,u,[Cc]$ by using equations of motion induced by superpotential: 
\begin{equation}
    \begin{aligned}
 & \frac{\partial W_{D_6}^{\prime}}{\partial [AC]} =D+C^*A^*=0,\quad &\frac{\partial W_{D_6}^{\prime}}{\partial D}=[AC]+ud=0,\\
 &  \frac{\partial W_{D_6}^{\prime}}{\partial [ac]}=d+c^*a^*=0,\quad  &\frac{\partial W_{D_6}^{\prime}}{\partial d}=[ac]+uD=0,\\
  &  \frac{\partial W_{D_6}^{\prime}}{\partial u}=[Cc]+Dd+v^2=0,\quad 
   & \frac{\partial W_{D_6}^{\prime}}{\partial [Cc]}=u+c^*C^*=0\,.
    \end{aligned}
\end{equation}

The final superpotential is 
\begin{equation}
\boxed{    W_{D_6}^{1}=-v^2c^*C^*-C^*A^*c^*a^*c^*C^*+[Aa]a^*A^*}\,.
\end{equation}

The $R$ charges of the dual theory are 
$R(C)=R(c)=R(A)=R(a)=\frac{1}{3}, R([Aa])=\frac{4}{3},R(v)=\frac{2}{3}$. The duality is verified by 
computing the Hilbert series and index of the dual theory.

\section{Conclusion}\label{conclusion}
In this paper, we performed a systematical study of the AdS/CFT pair, where the geometry side 
is given by the quotient singularity $\mathbb{C}^3/G$ with $G$ a finite subgroup of $SU(3)$ group. The field theory is given by the McKay correspondence, where the quiver is derived easily from the representation theory of $G$. Various physical properties such 
as quiver Hilbert series, and index, Seiberg duality, etc are computed from both the field theory 
and the gravity side. Our computation gives confirmation that the McKay quiver indeed gives one 
description of the SCFT on D3 branes probing the quotient singularity. The methods will be used to study 
 AdS/CFT pairs induced from other 3d Gorestein canonical singularities.

Many other aspects of AdS/CFT correspondence could be studied for those pairs, i.e. 
integrability \cite{zoubos2012review}, and non-conformal arrangement of the gauge group ranks \cite{Klebanov:2000hb}. Moreover, detailed KK studies on the 
gravity side are needed.

Among the field theories studied in this paper, those corresponding to finite subgroups of $SU(2)$ and $SO(3)$ groups are particularly interesting. In fact, the dual geometry belongs to the so-called compound du Val (cDV) singularity whose defining equation is given by $f_{ADE}(x,y,z)+wg(x,y,z,w)=0$. We would systematically study 
the AdS/CFT pair corresponding to those singularities in a separate publication.

\section*{Acklowledgement}
We would like to thank Wen-Bin Yan for helpful discussions.

\appendix
\section{Other cases }
Some other quiver gauge theories defined by finite subgroups of $SU(3)$ will be discussed in Appendix A like what we have done for finite groups of SO(3) in section \ref{fintesubgpso3}. Quiver gauge theories defined by Series E-L will be discussed in Appendix \ref{E-Lseries} and the branching rules for the irreducible representations of A series, $\Delta(3n^2)$ series, $\Delta(6n^2)$ series will be reviewed in Appendix \ref{Aseries}, \ref{delta3n2}, \ref{delta6n2}.

\subsection{$E, F, G, H, I, J, K, L$ series}\label{E-Lseries}
The coefficients of the tensor products of irreducible representations and equations of groups $E, F, G, H, I, J, K$ are summarized in \cite{gomi_nakamura_shinoda_2004, Yau1993GorensteinQS}.

We draw the McKay quivers and list the equations of $\mathbb{C}^3/G$ corresponding to groups $E, F, G, H, I, J, K, L$ series in Table \ref{EFGHIJKLseries}. The physical quantities computed from the quiver gauge theory are shown in Table \ref{physical E-L}.

\begin{table}[htbp]
    \begin{center}
 \begin{tabular}{|c|c|}
     \hline
     Group  $G$  &  Mckay quiver and the corresponding equation of $\mathbb{C}^3/G$ \\
     \hline
     $E$    & 
\begin{tikzcd}[row sep=tiny, column sep=tiny]
	&&& {1_1} \\
	&&& {1_0} \\
	&& {3_1} && {3_5} \\
	{3_6} &&&&&& {3_2} \\
	&&& {4_1/4_2} \\
	{3_3} &&&&&& {3_7} \\
	&& {3_8} && {3_4} \\
	&&& {1_3} \\
	&&& {1_2}
	\arrow[from=3-3, to=3-5]
	\arrow[from=3-3, to=4-1]
	\arrow[from=4-7, to=3-5]
	\arrow[from=6-1, to=4-1]
	\arrow[from=4-7, to=6-7]
	\arrow[from=6-1, to=7-3]
	\arrow[from=7-5, to=7-3]
	\arrow[from=7-5, to=6-7]
	\arrow[from=2-4, to=3-3]
	\arrow[from=3-5, to=2-4]
	\arrow[from=1-4, to=4-7]
	\arrow[from=4-1, to=1-4]
	\arrow[from=7-3, to=8-4]
	\arrow[from=8-4, to=7-5]
	\arrow[from=9-4, to=6-1]
	\arrow[from=6-7, to=9-4]
	\arrow[from=4-7, to=4-1]
	\arrow[from=3-3, to=7-3]
	\arrow[from=6-1, to=6-7]
	\arrow[from=7-5, to=3-5]
	\arrow[from=3-5, to=5-4]
	\arrow[from=4-1, to=5-4]
	\arrow[from=7-3, to=5-4]
	\arrow[from=6-7, to=5-4]
	\arrow[from=5-4, to=3-3]
	\arrow[from=5-4, to=4-7]
	\arrow[from=5-4, to=7-5]
	\arrow[from=5-4, to=6-1]
\end{tikzcd} \\
&$\left\{  \begin{array}{l}
9 y_{4}^{2}-12 y_{3}^{2}+y_{1}^{2} y_{3}-y_{1}^{2} y_{4}=0 \\
432 y_{5}^{2}-y_{2}^{3}+2y_{1}^{3}-36 y_{1} y_{4}+3 y_{1}^{2} y_{2}-36 y_{2} y_{3}=0 
\end{array}\right. $\\
\hline
F& \begin{tikzcd}[row sep=tiny, column sep=tiny]
	&&&& {6_2} \ar[llddddrr, rounded corners,to path= {
		-|  ([xshift=-30ex]\tikztotarget.west)
		|- (\tikztotarget)}]\\
	&& {3_5} & {3_6} && {3_7} & {3_8} \\
	{1_0} & {1_1} &&& 8 &&& {1_2} & {1_3} & 2 \ar[uulllll,rounded corners, to path= {
		|-  (\tikztotarget.east)}]\\
	&& {3_1} & {3_2} && {3_3} & {3_4} \\
	&&&& {6_1}\ar[rrrrruu,rounded corners, to path= {
		-|  (\tikztotarget.south)}]
	\arrow[from=3-1, to=4-3]
	\arrow[from=3-2, to=4-4]
	\arrow[from=4-3, to=2-3]
	\arrow[from=4-3, to=5-5]
	\arrow[from=4-4, to=2-4]
	\arrow[from=4-4, to=5-5]
	\arrow[from=4-6, to=2-6]
	\arrow[from=4-6, to=5-5]
	\arrow[from=4-7, to=2-7]
	\arrow[from=4-7, to=5-5]
	\arrow[from=2-3, to=3-5]
	\arrow[from=2-4, to=3-5]
	\arrow[from=2-6, to=3-5]
	\arrow[from=2-6, to=3-8]
	\arrow[from=2-7, to=3-5]
	\arrow[from=2-7, to=3-9]
	\arrow[from=1-5, to=2-3]
	\arrow[from=1-5, to=2-4]
	\arrow[from=1-5, to=2-6]
	\arrow[from=1-5, to=2-7]
	\arrow[from=3-5, to=4-3]
	\arrow[from=3-5, to=4-4]
	\arrow[from=3-5, to=4-6]
	\arrow[from=3-5, to=4-7]
	\arrow[ Rightarrow, from=3-5, to=1-5]
	\arrow[ Rightarrow, from=5-5, to=3-5]
	\arrow[from=2-3, to=3-1]
	\arrow[from=2-4, to=3-2]
	\arrow[from=3-8, to=4-6]
	\arrow[from=3-9, to=4-7]
\end{tikzcd}\\
&$
4 y_{4}^{3}-144 y_{2} y_{4}^{2}+1728 y_{2} y_{4} 
-\left(y_{1}^{3}-432 y_{3}^{2}-3 y_{1} y_{4}+36 y_{1} y_{2}\right)^{2}=0$
\\
\hline
G& \begin{tikzcd}[row sep=tiny, column sep=tiny]
	& {1_0} && {3_4} \\
	{2_2} & {3_1} && {8_1} & {6_4} & {2_3} \\
	{6_6} & {6_1} &&&& {6_2} & {3_2} & {1_1} \\
	&& {9_1} && {9_2} \\
	{3_6} & {8_3} && {3_7} && {8_2} & {3_5} \\
	{1_2} & {3_3} && {6_3} && {6_5} \\
	&&&&& {2_1}
	\arrow[from=3-2, to=2-4]
	\arrow[from=3-6, to=2-4]
	\arrow[from=3-2, to=5-2]
	\arrow[from=3-6, to=5-6]
	\arrow[from=6-4, to=5-2]
	\arrow[from=6-4, to=5-6]
	\arrow[from=5-4, to=4-3]
	\arrow[from=4-3, to=4-5]
	\arrow[from=4-5, to=5-4]
	\arrow[from=3-7, to=3-6]
	\arrow[from=3-7, to=5-7]
	\arrow[from=5-7, to=5-6]
	\arrow[from=3-8, to=3-7]
	\arrow[from=5-7, to=3-8]
	\arrow[from=5-1, to=5-2]
	\arrow[from=5-2, to=6-2]
	\arrow[from=6-2, to=5-1]
	\arrow[from=6-1, to=6-2]
	\arrow[from=5-1, to=6-1]
	\arrow[from=6-6, to=5-7]
	\arrow[from=6-6, to=6-4]
	\arrow[from=7-6, to=6-6]
	\arrow[from=6-4, to=7-6]
	\arrow[from=4-3, to=3-2]
	\arrow[from=4-3, to=3-6]
	\arrow[from=4-3, to=6-4]
	\arrow[from=2-4, to=4-3]
	\arrow[from=5-2, to=4-3]
	\arrow[from=5-6, to=4-3]
	\arrow[from=4-5, to=2-4]
	\arrow[from=4-5, to=5-6]
	\arrow[from=4-5, to=5-2]
	\arrow[from=6-6, to=4-5]
	\arrow[from=3-2, to=2-1]
	\arrow[from=2-1, to=3-1]
	\arrow[from=3-1, to=3-2]
	\arrow[from=3-1, to=4-5]
	\arrow[from=3-1, to=5-1]
	\arrow[from=2-2, to=1-4]
	\arrow[from=2-4, to=2-2]
	\arrow[from=2-2, to=3-2]
	\arrow[from=1-4, to=1-2]
	\arrow[from=1-2, to=2-2]
	\arrow[from=6-2, to=6-4]
	\arrow[from=1-4, to=2-4]
	\arrow[from=2-4, to=3-1]
	\arrow[from=5-6, to=3-7]
	\arrow[from=5-6, to=6-6]
	\arrow[from=5-2, to=6-6]
	\arrow[from=5-2, to=3-1]
	\arrow[from=2-4, to=2-5]
	\arrow[from=2-5, to=1-4]
	\arrow[from=2-5, to=3-6]
	\arrow[from=5-6, to=2-5]
	\arrow[from=2-5, to=4-5]
	\arrow[from=2-5, to=2-6]
	\arrow[from=3-6, to=2-6]
\end{tikzcd}\\
&$
4 y_{4}^{3}-9 y_{3} y_{4}^{2}+6 y_{3}^{2} y_{4}-2592 y_{1}^{2} y_{3} y_{4}-y_{3}^{3} 
+864 y_{1}^{2} y_{3}^{2}+6912 y_{2}^{3} y_{3}-186624 y_{1}^{4} y_{3}=0
$\\
\hline
H& 			\begin{tikzcd}[row sep=tiny, column sep=tiny]
				&[+20pt]&&[+20pt]&&3_2\ar[dl,bend left=10,""]\ar[dr,bend left=10,""]&\\[+30pt]
				1\ar[rr,bend left=10,""]&&3_1\ar[rr,bend left=10,""]\ar[ll,bend left=10,""]\ar[loop,out=290,in=250,looseness=12,""]&&5\ar[ll,bend left=10,""]\ar[rr,bend left=10,""]\ar[ur,bend left=10,""]\ar[loop,out=290,in=250,looseness=12,""]&&4\ar[ll,bend left=10,""]\ar[ul,bend left=10,""]\ar[loop,out=290,in=250,looseness=12,""]\\
			\end{tikzcd}\\
  & $
y_{4}^{2}+1728 y_{2}^{5}-y_{3}^{3}-720 y_{1} y_{2}^{3} y_{3} 
+80 y_{1}^{2} y_{2} y_{3}^{2}-64 y_{1}^{3}\left(5 y_{2}^{2}-y_{1} y_{3}\right)^{2}=0 
$\\
\hline
I&\begin{tikzcd}[row sep=0.7em, column sep=2.6em]
	& {3_2} & 8 \ar[loop, out=110,in=70,looseness=8]& \\
	{1_0} &&& 7\ar[loop, out=20,in=340,looseness=10]  \\
	& {3_1} & 6
	\arrow[from=3-2, to=1-2]
	\arrow[from=3-2, to=3-3]
	\arrow[from=1-2, to=2-1]
	\arrow[from=2-1, to=3-2]
	\arrow[from=1-2, to=1-3]
	\arrow[from=3-3, to=1-2]
	\arrow[curve={height=-6pt}, from=3-3, to=2-4]
	\arrow[curve={height=-6pt}, from=3-3, to=1-3]
	\arrow[curve={height=-6pt}, from=2-4, to=3-3]
	\arrow[curve={height=-6pt}, from=2-4, to=1-3]
	\arrow[from=1-3, to=3-2]
	\arrow[curve={height=-6pt}, from=1-3, to=3-3]
	\arrow[curve={height=-6pt}, from=1-3, to=2-4]
\end{tikzcd}\\
&$
y_{4}^{2}-y_{3}^{3}+88 y_{1}^{2} y_{2} y_{3}^{2}-1008 y_{1} y_{2}^{4} y_{3} 
-1088 y_{1}^{4} y_{2}^{2} y_{3}+256 y_{1}^{7} y_{3}-1728 y_{2}^{7} 
+60032 y_{1}^{3} y_{2}^{5}$\\
&
$-22016 y_{1}^{6} y_{2}^{3}+2048 y_{1}^{9} y_{2}=0 
$\\
\hline
\end{tabular}
\end{center}
\end{table}
\begin{table}[htbp]
\begin{center}
\begin{tabular}{|c|c|}
\hline
Group $G$ & Mckay quiver and the corresponding equation of $\mathbb{C}^3/G$\\
\hline

J&\begin{tikzcd}[row sep=small, column sep=small]
	{4_3} && {3_4} & {5_3} &&& {3_1} & {1_2} \\
	\\
	& {4_1} & {3_6} && {5_1} & {3_3} && {1_0} \\
	\\
	{4_2} && {3_5} & {5_2} &&& {3_2} & {1_1}
	\arrow[from=3-8, to=1-7]
	\arrow[from=5-8, to=5-7]
	\arrow[from=1-8, to=3-6]
	\arrow[from=1-7, to=1-8]
	\arrow[from=1-7, to=5-7]
	\arrow[from=5-7, to=3-8]
	\arrow[from=5-7, to=3-6]
	\arrow[from=3-6, to=5-8]
	\arrow[from=3-6, to=1-7]
	\arrow[from=1-3, to=1-1]
	\arrow[from=1-3, to=1-4]
	\arrow[from=5-3, to=3-2]
	\arrow[from=5-3, to=3-5]
	\arrow[from=3-3, to=5-1]
	\arrow[from=3-3, to=5-4]
	\arrow[from=3-2, to=1-3]
	\arrow[from=3-2, to=5-1]
	\arrow[from=3-2, to=5-4]
	\arrow[from=5-1, to=5-3]
	\arrow[from=5-1, to=1-1]
	\arrow[from=5-1, to=1-4]
	\arrow[from=1-1, to=3-2]
	\arrow[from=1-1, to=3-3]
	\arrow[from=1-1, to=3-5]
	\arrow[from=3-5, to=1-3]
	\arrow[from=3-5, to=1-7]
	\arrow[from=3-5, to=5-1]
	\arrow[from=3-5, to=5-4]
	\arrow[from=5-4, to=5-7]
	\arrow[from=5-4, to=5-3]
	\arrow[from=5-4, to=1-1]
	\arrow[from=5-4, to=1-4]
	\arrow[from=1-4, to=3-6]
	\arrow[from=1-4, to=3-3]
	\arrow[from=1-4, to=3-2]
	\arrow[from=1-4, to=3-5]
	\arrow[from=5-7, to=3-5]
	\arrow[from=3-6, to=5-4]
	\arrow[from=1-7, to=1-4]
\end{tikzcd}\\
&$
y_{4}^{3}-y_{3}\left[y_{2}^{2}+1728 y_{1}^{5}-720 y_{1}^{3} y_{4}
+80 y_{1} y_{4}^{2}-64 y_{3}\left(5 y_{1}^{2}-y_{4}\right)^{2}\right]=0 
$\\
\hline
K& \begin{tikzcd}[row sep=tiny, column sep=tiny]
	{1_0} & {3_4} & {8_1} &&&&& {7_3} \\
	& {3_1} &&& {6_3} \\
	& {3_5} \\
	{1_1} &&& {8_2} & {6_2} && {7_2} \\
	& {3_2} \\
	& {3_6} &&& {6_1} \\
	{1_2} & {3_3} & {8_3} &&&&& {7_1}
	\arrow[from=1-1, to=2-2]
	\arrow[from=4-1, to=5-2]
	\arrow[from=7-1, to=7-2]
	\arrow[from=2-2, to=1-2]
	\arrow[from=5-2, to=3-2]
	\arrow[from=7-2, to=6-2]
	\arrow[from=2-2, to=2-5]
	\arrow[from=5-2, to=6-5]
	\arrow[from=7-2, to=4-5]
	\arrow[from=1-2, to=1-1]
	\arrow[from=3-2, to=4-1]
	\arrow[from=6-2, to=7-1]
	\arrow[from=1-2, to=1-3]
	\arrow[from=3-2, to=4-4]
	\arrow[from=6-2, to=7-3]
	\arrow[from=6-5, to=6-2]
	\arrow[from=6-5, to=4-4]
	\arrow[from=6-5, to=4-7]
	\arrow[from=4-5, to=1-2]
	\arrow[from=4-5, to=1-8]
	\arrow[from=4-5, to=7-3]
	\arrow[from=2-5, to=1-3]
	\arrow[from=2-5, to=7-8]
	\arrow[from=2-5, to=3-2]
	\arrow[from=7-8, to=4-5]
	\arrow[from=7-8, to=4-7]
	\arrow[from=7-8, to=4-4]
	\arrow[from=4-7, to=2-5]
	\arrow[from=4-7, to=1-8]
	\arrow[from=1-8, to=6-5]
	\arrow[from=1-8, to=7-8]
	\arrow[from=1-8, to=1-3]
	\arrow[from=4-7, to=7-3]
	\arrow[from=1-3, to=2-2]
	\arrow[from=1-3, to=4-5]
	\arrow[from=1-3, to=4-7]
	\arrow[from=1-3, to=4-4]
	\arrow[from=4-4, to=5-2]
	\arrow[from=4-4, to=2-5]
	\arrow[from=4-4, to=1-8]
	\arrow[from=4-4, to=7-3]
	\arrow[from=7-3, to=7-2]
	\arrow[from=7-3, to=6-5]
	\arrow[from=7-3, to=7-8]
	\arrow[from=7-3, to=1-3]
\end{tikzcd}\\
&$
\begin{aligned}
&y_{4}^{3}-y_{3}(y_{2}^{2}+88 y_{3} y_{4}-1008 y_{1}^{4} y_{4}
-1088 y_{1}^{2} y_{3} y_{4}+256 y_{3}^{2} y_{4}\\&-1728 y_{1}^{7} +60032 y_{1}^{5} y_{3}
-22016 y_{1}^{3} y_{3}^{2}+2048 y_{1} y_{3}^{3} )=0 
\end{aligned}
$\\
\hline
L&\begin{tikzcd}[row sep=small, column sep=small]
	& {3_3} & {9_2} & {15_1} & {6_1} & {3_1} \\
	{5_1} & {8_2} & {9_1} & 10 & {8_1} & {1_0} & {5_2} \\
	& {3_4} & {9_3} & {15_2} & {6_2} & {3_2}
	\arrow[from=2-6, to=1-6]
	\arrow[curve={height=8pt}, from=1-6, to=3-6]
	\arrow[from=1-6, to=1-5]
	\arrow[from=3-6, to=2-6]
	\arrow[from=3-6, to=2-5]
	\arrow[from=1-2, to=1-3]
	\arrow[from=3-2, to=2-3]
	\arrow[curve={height=-50pt}, from=2-1, to=1-4]
	\arrow[curve={height=50pt}, from=2-7, to=1-4]
	\arrow[from=1-5, to=2-5]
	\arrow[from=1-5, to=2-4]
	\arrow[from=3-5, to=3-4]
	\arrow[from=2-5, to=1-6]
	\arrow[from=2-5, to=3-5]
	\arrow[from=2-5, to=1-4]
	\arrow[from=2-2, to=3-3]
	\arrow[from=2-2, to=1-4]
	\arrow[from=2-3, to=1-2]
	\arrow[from=2-3, to=3-3]
	\arrow[from=2-3, to=1-4]
	\arrow[from=1-3, to=2-2]
	\arrow[from=1-3, to=2-3]
	\arrow[from=1-3, to=2-4]
	\arrow[from=3-3, to=3-2]
	\arrow[from=3-3, to=3-4]
	\arrow[from=2-4, to=3-5]
	\arrow[from=2-4, to=3-3]
	\arrow[from=2-4, to=1-4]
	\arrow[from=1-4, to=1-5]
	\arrow[from=1-4, to=1-3]
	\arrow[curve={height=-50pt}, from=3-4, to=2-1]
	\arrow[curve={height=50pt}, from=3-4, to=2-7]
	\arrow[from=3-4, to=2-5]
	\arrow[from=3-4, to=2-2]
	\arrow[from=3-4, to=2-3]
	\arrow[from=3-4, to=2-4]
	\arrow[curve={height=8pt}, Rightarrow, from=1-4, to=3-4]
	\arrow[curve={height=-8pt}, from=3-3, to=1-3]
 	\arrow[from=3-5, to=3-6]
\end{tikzcd}\\
&$
\begin{aligned}
&459165024 y_{4}^{2}-25509168 y_{3}^{3}-(236196
+26244 \sqrt{15 \mathrm{i}}) y_{3}^{2} y_{1}^{5}+1889568(1+\sqrt{15 \mathrm{i}}) y_{3}^{2} y_{1}^{3} y_{2} \\
&+(8503056-2834352 \sqrt{15} \mathrm{i}) y_{3}^{2} y_{1} y_{2}^{2} 
-(891+243 \sqrt{15} \mathrm{i}) y_{3} y_{1}^{10}-(5346-8910 \sqrt{15} \mathrm{i}) y_{3} y_{1}^{8} y_{2} 
\\
&+(360612-51516 \sqrt{15} \mathrm{i}) y_{3} y_{1}^{6} y_{2}^{2}+(192456 +21384 \sqrt{15 \mathrm{i}}) y_{3} y_{1}^{4} y_{2}^{3}-3569184(1+\sqrt{15} \mathrm{i}) y_{3} y_{1}^{2} y_{2}^{4} \\
&-(7558272-2519424 \sqrt{15} \mathrm{i}) y_{3} y_{2}^{5} 
-2426112(1+\sqrt{15} \mathrm{i}) y_{2}^{7} y_{1}+(7978176
+886464 \sqrt{15} \mathrm{i}) y_{2}^{6} y_{1}^{3}\\
&-(3297168-471024 \sqrt{15} \mathrm{i}) y_{2}^{5} y_{1}^{5} 
+(78768-131280 \sqrt{15} \mathrm{i}) y_{2}^{4} y_{1}^{7}+(26928 
+7344 \sqrt{15} \mathrm{i}) y_{2}^{3} y_{1}^{9}\\
&-(1560-40 \sqrt{15} \mathrm{i}) y_{2}^{2} y_{1}^{11} 
+17(1-\sqrt{15} \mathrm{i}) y_{2} y_{1}^{13}=0
\end{aligned}
$\\
\hline
    \end{tabular}
    \caption{Mckay quivers and equations of $\mathbb{C}^3/G$ corresponding to groups $E, F, G, H, I, J, K, L$ series. }
    \label{EFGHIJKLseries}
    \end{center}
\end{table}

\begin{table}[htbp]
    \centering
    \begin{tabular}{|c|c|c|}
    \hline
    Group & Field theory data&\\
    \hline
    \multirow{4}{*}{$E$}&Central charges $(a,c)$ & $a=-\frac{21}{8}+27N^2$   \\ 
			&~&$ c=-\frac{7}{4}+27N^2$ \\ 
   \cline{2-3}
   &Quiver Hilbert series &  $H_{00}=\frac{(1-t^{12})(1-t^{16})}{(1-t^4)^2(1-t^6)(1-t^8)^2}$ \\ 
   \cline{2-3}
   &Single trace index &  $\mathcal{I}_{s.t.}=\frac{6t^6}{1-t^6}+\frac{2t^{24}}{(1-t^{24})}$ \\ 
\hline 
\multirow{4}{*}{$F$}&Central charges $(a,c)$ & $a=-3+54N^2$   \\ 
		&	~&$ c=-2+54N^2$ \\ 
  \cline{2-3}
		&	Quiver Hilbert series &  $H_{00}=\frac{(1-t^{24})}{(1-t^{4})(1-t^{6})(1-t^{8})(1-t^{8})}$ \\
\cline{2-3}
   &Single trace index &  $\mathcal{I}_{s.t.}=\frac{6t^6}{1-t^6}+\frac{3t^{24}}{(1-t^{24})}-\frac{t^{12}}{1-t^{12}}$ \\ 
  \hline
\multirow{4}{*}{$G$}&  Central charges $(a,c)$ & $a=-9/2+162N^2$   \\ 
&			~&$ c=-3+162N^2$ \\ \cline{2-3}
	&		Quiver Hilbert series &  $H_{00}=\frac{1-t^{36}}{(1-t^6)(1-t^8)(1-t^{12})^2}$ \\ 
 \cline{2-3}
   &Single trace index &  $\mathcal{I}_{s.t.}=\frac{4t^6}{1-t^6}+\frac{2t^{18}}{(1-t^{18})}+\frac{t^{24}}{(1-t^{24})}+\frac{2t^{36}}{(1-t^{36})}-\frac{t^{12}}{(1-t^{12})}$ \\ 
 \hline

\multirow{4}{*}{$ H$}&Central charges $(a,c)$ & 
	        $\begin{aligned}a=-1+ 15N^2
        \\ c=-\frac{3}{4} + 15N^2 \end{aligned}$ \\ 
	        \cline{2-3}
	        &Quiver Hilbert series &  $ H_{00}=\frac{1-t^{20}}{(1-t^{4/3})(1-t^4)(1-t^{20/3})(1-t^{10})}$ \\ 
          \cline{2-3}
   &Single trace index &  $\mathcal{I}_{s.t.}=\frac{4t^2}{1-t^2}+\frac{2t^{4}}{(1-t^{4})}+\frac{t^{14}}{(1-t^{14})}-\frac{3(t^2-t^4)}{(1-t^3y)(1-t^3y^{-1})}$ \\ 
	        \hline
   
\multirow{4}{*}{   $I$}& Central charges $(a,c)$ & $a=-\frac{7}{6}+42N^2$   \\ 
			&~&$ c=-\frac{5}{6}+42N^2$ \\ \cline{2-3}
			&Quiver Hilbert series &  $H_{00}=\frac{(1-t^{28})}{(1-t^{28/3})(1-t^{4})(1-t^{14})(1-t^{8/3})}$ \\ 
        \cline{2-3}
   &Single trace index &  $\mathcal{I}_{s.t.}=\frac{2t^2}{1-t^2}+\frac{t^{4}}{(1-t^{4})}+\frac{t^{6}}{(1-t^{6})}+\frac{t^{8}}{(1-t^{8})}+\frac{t^{14}}{(1-t^{14})}-\frac{2(t^2-t^4)}{(1-t^3y)(1-t^3y^{-1})}$ \\ 
   \hline
\multirow{4}{*}{  $J$}&Central charges $(a,c)$ & $a=-\frac{45}{16}+45N^2$   \\ 
	&		~&$ c=-\frac{15}{8}+45N^2$ \\ \cline{2-3}
	&		Quiver Hilbert series &  $H_{00}=\frac{1-t^{24}}{(1-t^4)^2(1-t^8)(1-t^{10})}$\\
  \cline{2-3}
   &Single trace index &  $\mathcal{I}_{s.t.}=\frac{6t^{6}}{(1-t^{6})}+\frac{2t^{12}}{(1-t^{12})}+\frac{t^{30}}{(1-t^{30})}$ \\ 
	 \hline
 \multirow{4}{*}{$K$}&Central charges $(a,c)$ & $a=-\frac{27}{8}+126N^2$   \\ 
	&		~&$ c=-\frac{9}{4}+126N^2$ \\ \cline{2-3}
	&		Quiver Hilbert series &  $H_{00}=\frac{1-t^{36}}{(1-t^4)(1-t^8)(1-t^{12})(1-t^{14})}$  \\
 \cline{2-3}
   &Single trace index &  $\mathcal{I}_{s.t.}=\frac{5t^{6}}{(1-t^{6})}+\frac{t^{12}}{(1-t^{12})}+\frac{t^{24}}{(1-t^{24})}+\frac{t^{42}}{(1-t^{42})}$ \\ 
 \hline
 \multirow{4}{*}{$L$}&Central charges $(a,c)$ & $a=-\frac{51}{16}+270N^2$   \\ 
		&	~&$ c=-\frac{17}{8}+270N^2$ \\ \cline{2-3}
		&	Quiver Hilbert series &  $H_{00}=\frac{(1-t^{60})}{(1-t^4)(1-t^8)(1-t^{20})(1-t^{30})}$ \\ 
  \cline{2-3}
   &Single trace index &  $\mathcal{I}_{s.t.}=\frac{6t^{6}}{(1-t^{6})}+\frac{t^{12}}{(1-t^{12})}+\frac{t^{24}}{(1-t^{24})}+\frac{t^{30}}{(1-t^{30})}$ \\ 
  \hline
    \end{tabular}
    \caption{central charges, Hilbert series and single trace index of series $E, F, G, H, I, J, K, L$.}
    \label{physical E-L}
\end{table}

\clearpage

\subsection{$A$ series: abelian subgroups}\label{Aseries}
The abelian groups are isomorphic to $Z_m\times Z_n$.
Set $g_{k,l}=\diag(\zeta_{m}^l, \zeta_n^k,\zeta_m^{-l}\zeta_n^{-k})$, with $\zeta_m, \zeta_n$ 
roots of unity with order $m$ and $n$. Define
\begin{equation*}
\rho_{ij}(g_{k,l})=\zeta_m^{ik} \zeta_n^{jl},~~i=1,\ldots, m,~~j=1,\ldots, n
\end{equation*}
which gives all the one dimensional representations of the group. Let $\pi=\rho_{1,0}\oplus \rho_{0,1}\oplus \rho_{-1,-1}$
be the natural representation s.t. $\pi(g_{k,l})=g_{k,l}$.
We have
\begin{equation*}
    \pi\otimes \rho_{ij}= \rho_{i+1,j} \oplus \rho_{i,j+1} \oplus \rho_{i-1,j-1}\,.
\end{equation*}
The quiver can be represented on a two-dimensional lattice drawn on the torus, which is the same 
as giving a dimer configuration.

\subsection{$\Delta(3n^2)$ series}\label{delta3n2}
Those quiver gauge theories were discussed in \cite{Muto:1998na}.
$\Delta(3n^2)$ is generated by $H_{n,n}\simeq \mathbb{Z}_n\otimes \mathbb{Z}_n$ and $T=\begin{pmatrix}
    0&1&0\\
    0&0&1\\
    1&0&0
\end{pmatrix}$.

For $n\in 3\mathbb Z$, there are 9 one-dimensional representations, and $\frac{n^2-3}{3}$ three-dimensional irreducible
representations. Denote the one-dimensional representations as $\theta_{0,0,m},\theta_{\frac{n}{3},\frac{2n}{3},m}, \theta_{\frac{2n}{3},\frac{n}{3},m}\,(m=1,2,3)$, and three-dimensional representations as $\theta_{i,j}$ where $(i,j) \in (1,2\ldots, n)\times (1,2,\ldots, n)\setminus\{(0,0),(\frac{n}{3},\frac{2n}{3}),(\frac{2n}{3},\frac{n}{3})\}$. The integers $i,j$ are defined modulo $n$, and there are equivalence relations among $\theta_{i,j}$:
\begin{equation}
   \theta_{i,j}=\theta_{-i+j,-i}=\theta_{-j,i-j}\,. 
\end{equation}

More explicitly, these representations are given by 
\begin{equation}\label{repre}
\begin{aligned}
     &\theta_{0,0,m}(T)=\xi_3^m\,,\quad \theta_{0,0,m}(g_{k,l})=1\,,\\
     &\theta_{\frac{n}{3},\frac{2n}{3},m}(T)=\xi_3^{m}\,,\quad \theta_{\frac{2n}{3},\frac{n}{3},m}(T)=\xi_3^{m}\,,
     \\
     &\theta_{\frac{n}{3},\frac{2n}{3},m}(g_{k,l})=\xi_3^{k+2l}\,,\quad \theta_{\frac{2n}{3},\frac{n}{3},m}(g_{k,l})=\xi_3^{2k+l}\,,
     \\
     &\theta_{i,j}(T)=T\,,\quad \theta_{i,j}(g_{k,l})=\diag(\xi_n^{ik+jl},\xi_n^{(j-i)k-il},\xi_n^{-jk+(i-j)l}),
\end{aligned}
\end{equation}
where $\xi_k=e^{2\pi i/k}$ and $g_{k,l}=\diag(\xi_n^k,\xi_n^l,\xi_n^{-k}\xi_n^{-l}) \in H_{n,n}$.

Then the natural representation $\pi$ can be given by 
\begin{equation}
\pi(g_{k,l})=\diag(\xi_m^k,\xi_m^l,\xi_m^{-k-l}),\pi(T)=T,
\end{equation}
and the decompositions of the tensor product are
\begin{align*}
   & \pi \otimes \theta_{0,0,m}=\theta_{0,1,m},~~ \pi \otimes \theta_{i,j}=\theta_{i-1,j-1}+\theta_{i,j+1}+\theta_{i+1,j}, \nonumber\\
   &     \pi \otimes \theta_{\frac{n}{3},\frac{2n}{3},m}=\theta_{\frac{n}{3}+1,\frac{2n}{3},m},~~   \pi \otimes \theta_{\frac{2n}{3},\frac{n}{3},m}=\theta_{\frac{2n}{3}+1,\frac{n}{3},m}\,.
\end{align*}

 For $n \not \in 3\mathbb Z$, there are 3 one-dimensional representations, and $\frac{n^2-1}{3}$ three-dimensional irreducible representations. The explicit representations are the same as (\ref{repre}) with the second line excluded. And the tensor product is also the same as before with the $\pi \otimes \theta_{\frac{n}{3},\frac{2n}{3},m}=\theta_{\frac{n}{3}+1,\frac{2n}{3},m}$ terms excluded.
 
The quiver diagrams for n=6 and 7 are shown in the Figure \ref{n=6,7}:
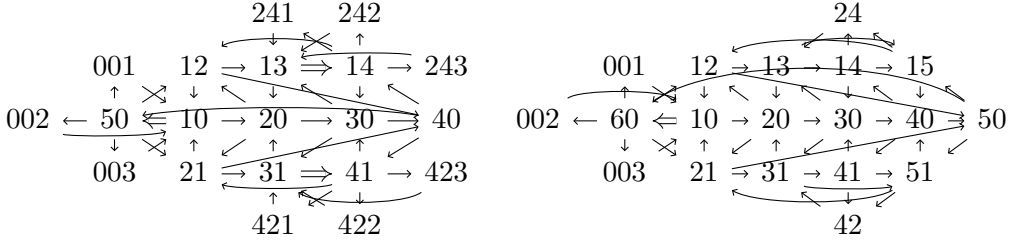
\begin{figure}[H]
    \centering
\begin{tikzcd}[row sep=tiny, column sep=tiny]
	&&& 241 & 242 \\
	& 001 & 12 & 13 & 14 & 243 \\
	002 & 50 & 10 & 20 & 30 & 40 \\
	& 003 & 21 & 31 & 41 & 423 \\
	&&& 421 & 422
	\arrow[from=2-3, to=2-4]
	\arrow[Rightarrow, from=2-4, to=2-5]
	\arrow[from=2-5, to=2-6]
	\arrow[from=2-3, to=3-3]
	\arrow[from=3-3, to=3-4]
	\arrow[from=2-4, to=3-4]
	\arrow[from=3-4, to=3-5]
	\arrow[from=2-5, to=3-5]
	\arrow[from=3-5, to=3-6]
	\arrow[from=3-6, to=2-5]
	\arrow[from=3-5, to=2-4]
	\arrow[from=3-4, to=2-3]
	\arrow[from=4-3, to=3-3]
	\arrow[from=4-3, to=4-4]
	\arrow[from=4-4, to=3-4]
	\arrow[Rightarrow, from=4-4, to=4-5]
	\arrow[from=4-5, to=3-5]
	\arrow[from=4-5, to=4-6]
	\arrow[from=3-4, to=4-3]
	\arrow[from=3-5, to=4-4]
	\arrow[from=3-6, to=4-5]
	\arrow[from=5-4, to=4-4]
	\arrow[from=4-5, to=5-4]
	\arrow[from=4-5, to=5-5]
	\arrow[from=5-5, to=4-4]
	\arrow[Rightarrow, from=3-3, to=3-2]
	\arrow[from=3-2, to=2-2]
	\arrow[from=2-2, to=3-3]
	\arrow[from=3-2, to=4-2]
	\arrow[from=4-2, to=3-3]
	\arrow[from=3-2, to=3-1]
	\arrow[from=2-3, to=3-6]
	\arrow[from=4-3, to=3-6]
	\arrow[from=1-4, to=2-4]
	\arrow[from=2-5, to=1-5]
	\arrow[from=2-5, to=1-4]
	\arrow[from=1-5, to=2-4]
	\arrow[curve={height=6pt}, from=2-6, to=2-4]
	\arrow[from=3-2, to=2-3]
	\arrow[curve={height=6pt}, from=3-1, to=3-3]
	\arrow[curve={height=12pt}, from=2-5, to=2-3]
	\arrow[curve={height=6pt}, from=3-6, to=3-2]
	\arrow[curve={height=-6pt}, from=4-5, to=4-3]
	\arrow[curve={height=-12pt}, from=4-6, to=4-4]
	\arrow[from=3-2, to=4-3]
\end{tikzcd}
\hspace{0.1em}
\begin{tikzcd}[row sep=tiny, column sep=tiny]
&&&& 24 \\
	& 001 & 12 & 13 & 14 & 15 \\
	002 & 60 & 10 & 20 & 30 & 40 & 50 \\
	& 003 & 21 & 31 & 41 & 51 \\
	&&&& 42
	\arrow[from=2-3, to=2-4]
	\arrow[from=2-4, to=2-5]
	\arrow[from=2-5, to=2-6]
	\arrow[from=2-3, to=3-3]
	\arrow[from=3-3, to=3-4]
	\arrow[from=2-4, to=3-4]
	\arrow[from=3-4, to=3-5]
	\arrow[from=3-5, to=3-6]
	\arrow[from=2-6, to=3-6]
	\arrow[from=2-5, to=3-5]
	\arrow[from=3-4, to=2-3]
	\arrow[from=3-5, to=2-4]
	\arrow[from=3-6, to=2-5]
	\arrow[from=4-3, to=3-3]
	\arrow[from=4-3, to=4-4]
	\arrow[from=4-4, to=3-4]
	\arrow[from=3-4, to=4-3]
	\arrow[from=3-5, to=4-4]
	\arrow[from=3-6, to=4-5]
	\arrow[from=4-4, to=4-5]
	\arrow[from=4-5, to=3-5]
	\arrow[from=4-6, to=3-6]
	\arrow[from=4-5, to=4-6]
	\arrow[from=3-6, to=3-7]
	\arrow[from=3-7, to=2-6]
	\arrow[from=3-7, to=4-6]
	\arrow[from=1-5, to=2-4]
	\arrow[from=2-6, to=1-5]
	\arrow[from=2-5, to=1-5]
	\arrow[from=4-5, to=5-5]
	\arrow[from=5-5, to=4-4]
	\arrow[from=4-6, to=5-5]
	\arrow[from=3-2, to=2-2]
	\arrow[from=3-2, to=3-1]
	\arrow[from=3-2, to=4-2]
	\arrow[from=3-2, to=4-3]
	\arrow[Rightarrow, from=3-3, to=3-2]
	\arrow[from=2-2, to=3-3]
	\arrow[from=4-2, to=3-3]
	\arrow[from=3-2, to=2-3]
	\arrow[curve={height=-12pt}, from=3-1, to=3-3]
	\arrow[curve={height=-12pt}, from=2-4, to=2-6]
	\arrow[curve={height=12pt}, from=2-6, to=2-3]
	\arrow[from=2-3, to=3-7]
	\arrow[from=4-3, to=3-7]
	\arrow[curve={height=6pt}, from=4-4, to=4-6]
	\arrow[curve={height=-12pt}, from=4-6, to=4-3]
	\arrow[curve={height=24pt}, from=3-7, to=3-2]
\end{tikzcd}
\caption{Quiver diagram for $\Delta(108)$(left) and $\Delta(147)$(right)}
\label{n=6,7}
\end{figure}

\subsection{$\Delta(6n^2)$ series}\label{delta6n2}
$\Delta(6n^2)$ is generated by $H_{n,n},T$ and $R=\begin{pmatrix}
  a&0&0\\
  0&0&b\\
  0&c&0
\end{pmatrix}$ where $abc=-1$.

If $3 \mid n$, there are 2 one-dimensional representations $\theta_{0,0,1},\theta_{0,0,2}$ , 4 two-dimensional representations  $\theta_{0,0,3},\theta_{\frac{m}{3},\frac{2m}{3},n_1} (n_1=1,2,3)$, $2(m-1)$ three-dimensional representations $\theta_{i,0,n_2} (n_2=1,2)$ and $\frac{m^2-2m}{6}$
six-dimensional irreducible representations $\theta_{i,j}$.

If $3\nmid n$, there are 2 one-dimensional representations $\theta_{0,0,1},\theta_{0,0,2}$, 1 two-dimensional $\theta_{0,0,3}$, $2(m-1)$ three-dimensional representation $\theta_{i,0,n_2} (n_2=1,2)$ and $\frac{m^2-3m+2}{6}$
six-dimensional irreducible representations $\theta_{i,j}$.

The natural representation is given by $\pi=\theta_{1,0,2}$, and the tensor products are

\begin{align*}
   & \pi \otimes \theta_{0,0,1}=\theta_{1,0,1}, \pi \otimes\theta_{0,0,2}=\theta_{1,0,2},\otimes\theta_{0,0,3}=\theta_{1,0,1}\oplus\theta_{1,0,2}\\
   &\pi\oplus \theta_{i,0,n_2}=\theta_{i+1,0,n_2}\oplus\theta_{-i,1-i}, \text{ for }i\neq 0, n_2=1,2. \\
   &   \pi \otimes \theta_{\frac{n}{3},\frac{2n}{3},2}=\theta_{\frac{n}{3}+1,\frac{2n}{3},2}, ( \pi \otimes \theta_{\frac{n}{3},\frac{2n}{3},1}=\theta_{\frac{n}{3}+1,\frac{2n}{3},1}, \pi \otimes \theta_{\frac{n}{3},\frac{2n}{3},3}=\theta_{\frac{n}{3}+1,\frac{2n}{3},3})\\
   &\pi \otimes \theta_{i,j}=\theta_{i-1,j-1}+\theta_{i,j+1}+\theta_{i+1,j}, \text{ for }i,j\notin \{0,\frac{m}{3},\frac{2m}{3}\}. 
\end{align*}

The relations in parentheses only appear in the case $3\mid n$.


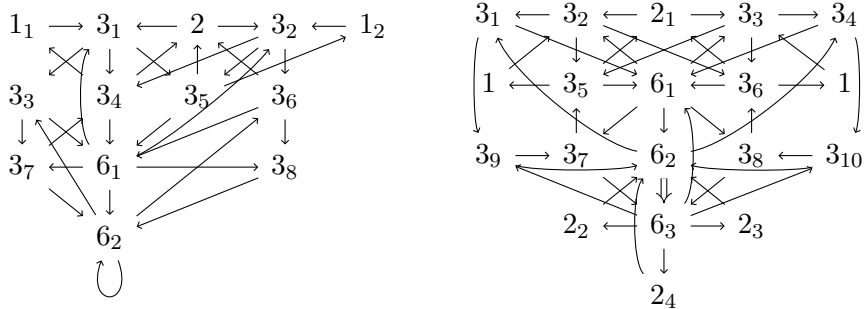
\begin{figure}[H]
    \centering
    \begin{tikzcd}[row sep=small, column sep=small]
	{1_1} & {3_1} & 2 & {3_2} & {1_2} \\
	{3_3} & {3_4} & {3_5} & {3_6} \\
	{3_7} & {6_1} && {3_8} \\
	& {6_2} \ar[loop, out=290,in=250,looseness=8]& \\
	\arrow[from=1-1, to=1-2]
	\arrow[from=1-3, to=1-2]
	\arrow[from=1-3, to=1-4]
	\arrow[from=1-5, to=1-4]
	\arrow[from=1-2, to=2-2]
	\arrow[from=1-2, to=2-1]
	\arrow[from=1-4, to=2-2]
	\arrow[from=2-2, to=1-3]
	\arrow[from=2-3, to=1-3]
	\arrow[from=1-4, to=2-3]
	\arrow[from=1-4, to=2-4]
	\arrow[from=2-2, to=1-1]
	\arrow[from=1-2, to=2-3]
	\arrow[from=2-4, to=1-3]
	\arrow[from=2-1, to=3-1]
	\arrow[from=3-1, to=2-2]
	\arrow[from=2-2, to=3-2]
	\arrow[from=3-2, to=3-4]
	\arrow[from=2-4, to=3-4]
	\arrow[from=3-2, to=4-2]
	\arrow[from=3-2, to=3-1]
	\arrow[from=2-1, to=3-2]
	\arrow[from=4-2, to=2-1]
	\arrow[from=3-1, to=4-2]
	\arrow[from=3-4, to=4-2]
	\arrow[from=4-2, to=2-4]
	\arrow[from=2-3, to=3-2]
	\arrow[from=2-4, to=3-2]
	\arrow[curve={height=-12pt}, from=3-2, to=1-2]
	\arrow[curve={height=6pt}, from=3-2, to=1-4]
	\arrow[from=2-3, to=1-5]
\end{tikzcd}
\hspace{1.5em}
\begin{tikzcd}[row sep=small, column sep=small]
	{3_1} & {3_2} & {2_1} & {3_3} & {3_4} \\
	1 & {3_5} & {6_1} & {3_6} & 1 \\
	{3_9} & {3_7} & {6_2} & {3_8} & {3_{10}} \\
	& {2_2} & {6_3} & {2_3} \\
	&& {2_4}
	\arrow[from=1-2, to=1-1]
	\arrow[from=1-3, to=1-2]
	\arrow[from=1-3, to=1-4]
	\arrow[from=1-4, to=1-5]
	\arrow[from=2-1, to=1-2]
	\arrow[from=2-2, to=2-1]
	\arrow[from=1-2, to=2-2]
	\arrow[from=2-2, to=2-3]
	\arrow[from=2-2, to=1-3]
	\arrow[from=2-4, to=2-3]
	\arrow[from=2-3, to=1-4]
	\arrow[from=2-5, to=1-4]
	\arrow[from=2-4, to=2-5]
	\arrow[from=1-4, to=2-4]
	\arrow[from=2-4, to=1-3]
	\arrow[from=2-3, to=1-2]
	\arrow[from=1-1, to=2-3]
	\arrow[from=1-5, to=2-3]
	\arrow[from=1-4, to=2-2]
	\arrow[from=1-2, to=2-4]
	\arrow[from=2-3, to=3-3]
	\arrow[curve={height=-12pt}, from=3-3, to=1-1]
	\arrow[curve={height=12pt}, from=3-3, to=1-5]
	\arrow[from=3-2, to=2-2]
	\arrow[from=3-1, to=3-2]
	\arrow[from=3-2, to=4-3]
	\arrow[from=2-3, to=3-2]
	\arrow[curve={height=6pt}, from=1-1, to=3-1]
	\arrow[from=4-3, to=3-1]
	\arrow[curve={height=6pt}, from=3-1, to=3-3]
	\arrow[Rightarrow, from=3-3, to=4-3]
	\arrow[from=4-2, to=3-3]
	\arrow[from=4-3, to=4-2]
	\arrow[from=4-3, to=5-3]
	\arrow[from=4-3, to=4-4]
	\arrow[from=3-4, to=4-3]
	\arrow[from=3-5, to=3-4]
	\arrow[from=2-3, to=3-4]
	\arrow[from=3-4, to=2-4]
	\arrow[curve={height=-6pt}, from=1-5, to=3-5]
	\arrow[from=4-3, to=3-5]
	\arrow[from=4-4, to=3-3]
	\arrow[curve={height=-12pt}, from=5-3, to=3-3]
	\arrow[curve={height=12pt}, from=4-3, to=2-3]
	\arrow[curve={height=-6pt}, from=3-5, to=3-3]
\end{tikzcd}
    \caption{Mckay quivers of the $\Delta(150)$(Left) and $\Delta(216)$(Right)}
    \label{n=5,6}
\end{figure}

\bibliographystyle{JHEP}

\bibliography{ref}

\end{document}